\documentclass[12pt,a4paper]{article}
\pdfoutput=1

\usepackage{amsmath,amssymb,bm,multirow,bbold,hyperref,slashed}
\usepackage{graphicx}
\usepackage[american]{babel}
\usepackage{latexsym}
\usepackage{mathrsfs}
\usepackage{amsfonts}
\usepackage{dcolumn} 
\usepackage{float}
\usepackage{cite}

\usepackage{amsmath}
 \usepackage{feynmf}

\usepackage[utf8]{inputenc}
\usepackage{epsfig}

\newcommand{\lra}[1]{\overset{\text{\tiny$\bm\leftrightarrow$}}{#1}}
\newcommand{\la}[1]{\overset{\text{\tiny$\bm\leftarrow$}}{#1}}
\allowdisplaybreaks
\begin{document}

\begin{titlepage}
\begin{flushright}
{
IFIC/16-28\\
TUM-HEP-1047/16
}
\end{flushright}
\vskip 1cm

\setcounter{footnote}{0}
\renewcommand{\thefootnote}{\fnsymbol{footnote}}
\vspace*{1cm} 
\begin{center}
{\LARGE \bf Integrating out heavy particles with }
\\ [13pt]{\LARGE \bf functional methods: a simplified framework}
\vspace{2cm} \\
{\sc  Javier~Fuentes-Mart\'{\i}n}$^{1,}$\footnote{Email:~Javier.Fuentes@ific.uv.es}
{\sc , Jorge~Portol\'es}$^{1,}$\footnote{Email:~Jorge.Portoles@ific.uv.es} and
{\sc Pedro~Ruiz-Femen\'ia}$^{2,}$\footnote{Email:~Pedro.Ruiz-Femenia@tum.de}
\vspace{1.5cm} \\
$^1$Instituto de F\'{\i}sica Corpuscular, CSIC - Universitat de Val\`encia, 
Apt. Correus 22085, E-46071 Val\`encia, Spain \\ [13pt]
$^2$Physik Department T31, James-Franck-Stra\ss e, Technische Universit\"at M\"unchen, D–85748 Garching, Germany
\end{center}

\setcounter{footnote}{0}
\renewcommand{\thefootnote}{\arabic{footnote}}
\vspace*{1cm}

\date{\today}

\begin{abstract}
We present a systematic procedure to obtain the one-loop low-energy effective Lagrangian resulting from integrating out the heavy fields of a given ultraviolet theory. We show that the matching coefficients are determined entirely by the hard region of the functional determinant involving the heavy fields. This represents an important 
simplification with respect the conventional matching approach, where the full and effective theory contributions have to be computed separately and a cancellation of the infrared divergent parts has to take place. We illustrate the method with a descriptive toy model and with an extension of the Standard Model with a heavy 
real scalar triplet. A comparison with other schemes that have been put forward recently is also provided.
\end{abstract}
\end{titlepage}

\section{Introduction}\label{s:intro}
New physics searches at the LHC rely, namely, in the discovery of a new spectrum of particles with masses much 
larger than the electroweak scale though it is being seen that they can be rather elusive. Our present understanding of the laws of physics tells us that whether these are supersymmetric states or an extended scalar sector, for instance, their role at the electroweak scale should be weighted by inverse powers of their masses. This is the main tenet behind our concept and 
compelling use of effective field theories in particle physics: we obtain the low-energy theory by integrating out the heavier spectrum in the, up to now 
model-dependent, ultraviolet completion of the former. In this way we determine the marks of the underlying theory at higher scales on the low-energy couplings, {\em i.e.} Wilson coefficients, of the effective field theory (EFT). Upon comparison with the electroweak scale phenomenology we should be able to obtain information on new physics scenarios. This framework has pervaded the last fifty years of research in particle physics.
\par 
Although the rationale and the procedure has been well developed long ago in the literature (see for instance  \cite{Georgi:1985kw,Donoghue:1992dd}), the integration at 
next-to-leading order in the upper theory, that is to say at one loop, is undergoing lately an intense debate~\cite{Henning:2014wua,Drozd:2015rsp,delAguila:2016zcb,Boggia:2016asg,Henning:2016lyp,Ellis:2016enq} that, as we put forward in this paper, still allows for simpler
alternatives. There are two techniques to obtain the Wilson coefficients of the EFT. The most employed one amounts to matching the diagrammatic computation of
given Green Functions with light particle external legs 
in the full theory, where heavy states can appear in virtual lines, and in the EFT, at energies where the EFT can describe the dynamics of the light particles as an expansion in inverse powers of the heavy particle mass scale. 
Alternatively one can perform the functional integration of the heavier states without being concerned with specific Green Functions, and later extract the local contributions 
that are relevant for the description of the low-energy dynamics of the light fields. This last methodology was applied, for example, 
in Refs.~\cite{Dittmaier:1995cr,Dittmaier:1995ee}, to obtain the non-decoupling effects of a heavy Higgs in the Standard Model (SM).
The path integral formulation has obvious advantages over the matching procedure as, for instance, one does not need to handle Feynman diagrams
nor symmetry factors, and one obtains directly the whole set of EFT operators together
with their matching conditions, {\em i.e.} no prior knowledge about the specifics of the EFT operator structure, symmetries, etc., is required.
\par 
One of the issues recently arisen involves the widely used technique to perform the functional integration set up more than thirty years ago by
the works of Aitchison and Fraser~\cite{Fraser:1984zb,Aitchison:1984ys,Aitchison:1985pp,Aitchison:1985hu}, Chan~\cite{Chan:1985ny,Chan:1986jq}, Gaillard~\cite{Gaillard:1985uh} and Cheyette~\cite{Cheyette:1985ue}. As implemented by
 Refs.~\cite{Henning:2014wua,Drozd:2015rsp}, this technique did not include all the one-loop contributions from the integration, in particular those where heavy and light field quantum fluctuations appear in the same loop. This fact was noticed in Ref.~\cite{delAguila:2016zcb}, and
 fixed later on in Refs.~\cite{Henning:2016lyp,Ellis:2016enq}, by the use of variants of
 the functional approach which require additional ingredients in order to subtract the
 parts of the heavy-light loops which are already accounted for by the one-loop EFT
 contribution. Here we would like to introduce a 
 more direct method to obtain the one-loop effective theory that builds upon
 the works of Refs.~\cite{Dittmaier:1995cr,Dittmaier:1995ee}, 
 and that uses the technique of  \lq \lq expansion by regions" \cite{Beneke:1997zp,Smirnov:2002pj,Jantzen:2011nz} to read off the one-loop matching coefficients from the 
 full theory computation, thus bypassing the need of subtracting any infrared contribution.
 In short, the determination of the one-loop EFT in the approach we propose reduces to the calculation of the {\em hard} 
 part of the determinant of $\widetilde{\Delta}_H$, where $\widetilde{\Delta}_H$ arises from the diagonalization of the quadratic term in the expansion of the full theory Lagrangian around the classical field configurations, and the determinant is just the result of the Gaussian integration
 over the heavy quantum fluctuations. In this way, the terms that mix light and heavy spectra inside the loop get disentangled by means of  
a field transformation in the path integral
 that brings the quadratic fluctuation into diagonal form: The part involving only the light quantum fields remains untouched by the transformation and all heavy particle effects
 in the loops are shifted to the modified heavy quadratic form $\widetilde{\Delta}_H$.
This provides a  conceptually simple and straightforward technique to obtain all the 
one-loop local EFT couplings from an underlying theory that can contain arbitrary interactions
between the heavy and the light degrees of freedom.
\par 
The contents of the paper are the following. The general outline of the method is given in Section~\ref{s:method}, 
where we describe the transformation that diagonalizes the quadratic fluctuation which defines $\widetilde{\Delta}_H$, 
and then discuss how to extract the contributions from $\widetilde{\Delta}_H$ that are relevant for determining the one-loop EFT. 
In Section~\ref{s:murayama} we compare our procedure with those proposed recently 
by \cite{Henning:2014wua,Henning:2016lyp} and \cite{Drozd:2015rsp,Ellis:2016enq}.
The virtues of our method are better seen through examples: first 
we consider a simple scalar toy model in Section~\ref{s:examples}, where we can easily illustrate the advantages of our procedure
with respect the conventional matching approach; then we turn to an extension of the SM with a
heavy real scalar triplet, that has been used as an example in recent papers. We conclude with
Section~\ref{s:conclusions}. Additional material concerning the general formulae for dimension-six operators, and the expression of the fluctuation operator in the SM case is provided in the appendices.

\section{The method}\label{s:method}

We outline in this section the functional method to determine the EFT Lagrangian 
describing the dynamics of light particles at energies much smaller than $m_H$, the typical mass of a heavy particle, or set of particles, 
that reproduces the full-theory results at the one-loop level.
The application of the method
to specific examples is postponed to Section~\ref{s:examples}.

Let us consider a general theory whose field content can be split into heavy ($\eta_H$) and light ($\eta_L$) degrees of freedom, that we 
collect generically in $\eta=\left(\eta_H\,,\eta_L\right)$. 
For charged degrees of freedom, the field and its complex conjugate enter as separate components in $\eta_H$ and $\eta_L$. In order to obtain the one-loop effective action, we split each field component into a background field configuration, $\hat{\eta}$, which satisfy the classical equations of motion (EOM), and 
a quantum fluctuation $\eta$, 
{\em i.e.} we write $\eta \to \hat{\eta} + \eta$. Diagrammatically, the background part corresponds to tree lines in Feynman graphs while lines inside loops arise from the quantum fields; this means that terms higher than quadratic in
the quantum fields yield vertices that can only appear in diagrams at higher loop orders. Therefore, at the one-loop level one has to consider only the Lagrangian up to terms 
quadratic in $\eta$: 
\begin{align}\label{eq:BFM}
\mathcal{L}=\mathcal{L}^{\mbox{\scriptsize tree}}(\hat{\eta})+\mathcal{L}^{(\eta^2)}+\mathcal{O}\left(\eta^3\right)\,.
\end{align}
The zeroth order term, $\mathcal{L}^{\mbox{\scriptsize tree}}$, depends only on the classical field configurations and yields the tree-level effective action. At energies much lower than the mass of the heavy fields, the background heavy fields $\hat{\eta}_H$ can be eliminated from the tree-level action by using their EOM.
The linear term in the expansion of $\mathcal{L}$ around the background fields is, up to a total derivative, proportional to the EOM evaluated at $\eta=\hat{\eta}$, and thus vanishes.  
From the  quadratic piece 
\begin{align}\label{eq:2ndorderL} 
\mathcal{L}^{(\eta^2)}=\frac{1}{2} \eta^\dagger
\frac{\partial^2\mathcal{L}}{\partial\eta^*\,\partial\eta}\bigg|_{\eta=\hat\eta}
\; \eta \equiv \frac{1}{2}\eta^\dagger\,\mathcal{O}\,\eta\,,
\end{align}
we identify the fluctuation operator $\mathcal{O}$, with generic form
\begin{align}\label{eq:fluctuation_op}
\mathcal{O}=
\begin{pmatrix}
\Delta_H & X_{LH}^\dagger\\
X_{LH}   & \Delta_L
\end{pmatrix}\,,
\end{align}
and which depends only on the classical fields $\hat\eta$.

The one-loop effective action thus derives from the path integral 
\begin{align}\label{eq:1LEFT_def_ini}
e^{iS}
={\cal N}\int\mathcal{D}\eta_L\mathcal{D}\eta_H\, \exp \left[ i\int dx\,\mathcal{L}^{(\eta^2)} \right]
\,,
\end{align}
which can be obtained by Gaussian integration. Our aim is to compute the one-loop heavy particle effects 
in the Green functions of the light fields as an expansion in the heavy mass scale $m_H$.  
In terms of Feynman diagrams, the latter corresponds to computing all one-loop diagrams
involving heavy lines and expanding them in $1/m_H$. 
The latter can be formally 
achieved by doing the functional integration over the fields $\eta_H$.
However, 
the presence of mixing terms among heavy and light quantum fields in $\mathcal{L}^{(\eta^2)}$ (equivalently, of one-loop
diagrams with both heavy and light lines inside the loop), 
makes it necessary to first
rewrite the fluctuation operator in Eq.~\eqref{eq:fluctuation_op} in an equivalent block-diagonal form.
A way of achieving this is by performing shifts (with unit Jacobian determinant) in the quantum fields, which 
can be done in different ways. We choose  
a field transformation that shifts the information about the mixing terms $X_{LH}$
in the fluctuation operator into a redefinition of the heavy-particle block $\Delta_H$, while leaving $\Delta_L$
untouched. This has the advantage that all heavy particle effects in the one-loop effective action are thus obtained
through the computation of the determinant that results from the path integral over the heavy fields. 
This shifting  procedure was actually used in Refs.~\cite{Dittmaier:1995cr,Dittmaier:1995ee} 
for integrating out the Higgs field in the SU(2)
gauge theory and in the SM.
An alternative shift, which is implicitly used in Ref.~\cite{Henning:2016lyp}, will be discussed in Section~\ref{s:murayama}. 

The explicit form of the field transformation that brings $\mathcal{O}$ into the desired block-diagonal form reads
\begin{align}\label{eq:Pdef}
P=
\begin{pmatrix}
I & 0\\
-\Delta_L^{-1}X_{LH} & I
\end{pmatrix}\,,
\end{align}
and one immediately obtains
\begin{align}\label{eq:field_redef}
P^\dagger \mathcal{O}P=
\begin{pmatrix}
\widetilde{\Delta}_H & 0\\
0   & \Delta_L
\end{pmatrix}\,,
\end{align}
with 
\begin{align}\label{eq:tildeDeltaH}
\widetilde{\Delta}_H = \Delta_H-X_{LH}^\dagger\Delta_L^{-1}X_{LH}\,.
\end{align}

The functional integration over the heavy fields $\eta_H$ can now be carried out easily,
\begin{align}\label{eq:1LEFT_def}
e^{iS}
=
 \left( \det \widetilde{\Delta}_H \right)^{-c} \,
{\cal N}\int\mathcal{D}\eta_L\, \exp \left[ i\int dx\;\frac{1}{2}\eta_L^\dagger\Delta_L \eta_L \right]
\, ,
\end{align}
with $c=1/2,-1$ depending on the bosonic or fermionic nature of the heavy fields. For simplicity, we assume 
that all degrees of freedom in the heavy sector are either bosons or fermions. In the case of mixed statistics, 
one needs to further diagonalize $\widetilde{\Delta}_H$ to decouple the bosonic and fermionic blocks.
The remaining Gaussian integration in Eq.~\eqref{eq:1LEFT_def} reproduces the one-loop contributions with light particles
running inside the loop, and 
heavy fields can appear only as tree-level lines through the dependence of $\Delta_L$
in $\hat{\eta}_H$. 
We thus define the part of the one-loop effective action coming from loops involving heavy fields as
\begin{align}
S_{H}&=i\,c\,\ln\det\widetilde{\Delta}_H\,.
\end{align}
In order to compute the determinant of $\widetilde{\Delta}_H$ we use standard techniques developed in
the literature~\cite{Chan:1985ny,Ball:1988xg}. First it is rewritten as
\begin{align}
S_{H} &= i\,c\,\mbox{Tr}\ln\widetilde{\Delta}_H ,
\end{align}
where $\mbox{Tr}$ denotes the full trace of the operator, also in coordinate space. It is convenient for our purposes
to rewrite the functional trace using momentum eigenstates defined in $d$ dimensions as
\begin{align}\label{eq:functrace}
\begin{aligned}
S_{ H} &=
i\,c\,\mbox{tr}\int \frac{d^dp}{\left(2\pi\right)^d}\; \langle\, p|\ln \widetilde{\Delta}_H | p\rangle\\
&=i\,c\,\mbox{tr}\int d^dx\int \frac{d^dp}{\left(2\pi\right)^d}\; e^{-ipx}\ln\Big(\widetilde{\Delta}_H\left(x,\partial_x\right)\Big)\,e^{ipx}\\
&=i\,c\,\mbox{tr}\int d^dx\int \frac{d^dp}{\left(2\pi\right)^d}\; \ln\Big( \widetilde{\Delta}_H \left(x,\partial_x+ip\right)\Big)\, \mathbb{1}
\,.
\end{aligned}
\end{align}
The derivatives in $\widetilde{\Delta}_H$ yields factors of $ip$ upon acting on the 
exponentials\footnote{
Note that $\widetilde{\Delta}_H$ can also depend in $\partial_x^\intercal$.
Transpose derivatives are defined from 
the adjoint operator, which acts on the function at the left, and can 
be replaced by $-\partial_x$, the difference being a total derivative term. The identity
$\mathbb{1}$ in Eq.~(\ref{eq:functrace}) serves as a reminder that derivatives at the rightmost disappear
after acting on the exponential.
}. The symbol $\mbox{tr}$ denotes the trace over internal degrees of freedom only.
Since $\widetilde{\Delta}_H$ contains the kinetic term of the heavy fields, in the case of 
scalar fields it has the generic form
\begin{align} \label{eq:Udefined}
\widetilde{\Delta}_H = -\hat{D}^2-m_H^2-U\,,
\end{align}
with $\hat D_\mu$ denoting the covariant derivative for the heavy fields with
background gauge fields. Performing the shift $\partial_x\to\partial_x+ip$ we find 
\begin{align}\label{eq:Leff}
\begin{aligned}
S_{H}&= \frac{i}{2}\,\mbox{tr} \int d^dx \int \frac{d^dp}{\left(2\pi\right)^d}\; 
\ln\left(p^2-m_H^2- 2ip\hat{D}-\hat{D}^2-U\left(x,\partial_x+ip\right)\right) \mathbb{1}
\,.
\end{aligned}
\end{align}
For fermions, the same formula, Eq.~(\ref{eq:Leff}), applies but with an overall minus sign 
 and with $U$ replaced
by
\begin{align}\label{eq:Uferm}
U_{\rm{ferm.}}=-\frac{i}{2} \sigma^{\mu\nu}\left[\hat{D}_\mu,\hat{D}_\nu\right]-i\left[\hat{\slashed{D}},\Sigma_e\right]+i\left\{\hat{\slashed{D}},\Sigma_o\right\}+2m_H\Sigma_e+\Sigma\left(\Sigma_e-\Sigma_o\right)\,.
\end{align}
Here $\Sigma\equiv\Sigma_e+\Sigma_o$ is defined by $\widetilde{\Delta}_H=i\hat{\slashed{D}}-m_H-\Sigma$, and $\Sigma_e$ ($\Sigma_o$) contains an even (odd) number of gamma matrices. 
Finally, we can Taylor expand the logarithm to get
\begin{align}\label{eq:L1loop_final}
S_{H}=\mp\frac{i}{2} \int d^dx\,\sum_{n=1}^\infty\frac{1}{n}\int\frac{d^dp}{\left(2\pi\right)^d}\,\mbox{tr}\left\{\left(\frac{2ip\hat{D}+\hat{D}^2+U\left(x,\partial_x+ip\right)}{p^2-m_H^2}\right)^n  \mathbb{1} \right\}\, \,,
\end{align}
where we have dropped an irrelevant constant term, and the negative (positive) global sign corresponds to the integration of boson (fermion) heavy fields.

The effective action Eq.~(\ref{eq:L1loop_final}) generates all one-loop amplitudes with at least one heavy particle propagator in the loop. 
One-loop diagrams with $n$ heavy propagators are reproduced from the $n$-th term in the expansion of Eq.~(\ref{eq:L1loop_final}). In addition the diagram can contain light propagators, that arise upon  expanding the term 
$X^\dagger_{LH}\Delta_L^{-1}X_{LH}$ in $\widetilde \Delta_H$ using
\begin{align}\label{eq:DeltaL}
\Delta_L^{-1}=\sum_{n=0}^\infty\left(-1\right)^n\,\left(\widetilde \Delta_L^{-1}X_L\right)^n\widetilde \Delta_L^{-1}\,,
\end{align}
which corresponds to the Neumann series expansion of $\Delta_L^{-1}$, and we have made the separation $\Delta_L=\widetilde \Delta_L+X_L$, with $\widetilde \Delta_L$ corresponding to the the fluctuations coming from the kinetic terms, {\em i.e.} $\widetilde \Delta_L^{-1}$ is the light field propagator. From the definition of the fluctuation operator ${\mathcal O}$, Eq.~(\ref{eq:fluctuation_op}), the terms in $\widetilde \Delta_L$ are part of the diagonal components
of ${\mathcal O}$. At the practical level, for the calculation of $\Delta_L^{-1}$ using Eq.~(\ref{eq:DeltaL}) it is simpler to define $\widetilde \Delta_L$ directly as the whole diagonal of ${\mathcal O}$.

Loops with heavy particles receive contributions from the region of {\it hard} loop momenta   $p\sim m_H$, and from the {\it soft} momentum region, where the latter is set by the low-energy scales in the theory, either $p \sim m_L$ or any of the light-particle external momenta, $p_i\ll m_H$. In dimensional regularization the two contributions can be computed separately by using the so-called ``expansion by regions"~\cite{Beneke:1997zp,Smirnov:2002pj,Jantzen:2011nz}. In this method the contribution of each region is obtained by
expanding the integrand into a Taylor series with respect to the
parameters that are small there, and then integrating every region over the
full $d$-dimensional space
of the loop momenta. In the hard region, all the low-energy scales are expanded out and  only $m_H$ remains in the propagators. The resulting integrand yields local contributions in the form of a polynomial in the low-energy momenta and masses, with factors of $1/m_H$ to adjust the dimensions. This part is therefore fully determined by the short-distance behaviour of the full theory and has to be included into the EFT Lagrangian in order to match the amplitudes in the full and effective theories. Indeed, the coefficients of the polynomial terms from the hard contribution of a given (renormalized) amplitude provide the one-loop matching coefficients of corresponding local terms in the effective theory. This can be understood easily since the soft part of the amplitude results upon expanding the vertices and propagators according to $p\sim m_L \ll m_H$, with $p$ the loop momentum. This expansion, together with the one-loop terms with light particles that arise from the Gaussian integral of $\Delta_L$ in Eq.~(\ref{eq:1LEFT_def}),
yields the same one-loop amplitude as one would obtain using the Feynman rules of the effective Lagrangian for the light fields obtained by tree-level matching, equivalently the Feynman rules from ${\cal L}^{\rm tree}$ in Eq.~(\ref{eq:BFM}) where the background heavy field $\hat{\eta}_H$ has been eliminated in favour of $\hat{\eta}_L$ using the classical EOM. 
Therefore, in the difference of the full-theory and EFT renormalized amplitudes at one-loop only the hard part of the full-theory amplitude remains, and one can read off the one-loop matching coefficients directly from the computation of the latter. Let us finally note that in the conventional matching approach, the same infrared regularization has to be used in the full and EFT calculations, in order to guarantee that the infrared behaviour of both theories is identical.
This is of course fulfilled in the approach suggested here, since the one-loop EFT amplitude is defined implicitly by the full theory result. Likewise, the ultraviolet (UV) divergences of the EFT are determined by UV divergences in the soft part, that are regulated in $d$ dimensions in our approach. For the renormalization of the amplitudes, we shall use the $\overline{\rm MS}$ subtraction scheme. 

Translated into the functional approach, the preceding discussion implies that 
the EFT Lagrangian  at one-loop is then determined as 
\begin{align}\label{eq:S_Hhard}
\int d^dx \, {\cal L}^{\rm 1loop}_{\mbox{\tiny{EFT}}} = S_{H}^{\rm hard}
\,,
\end{align}
where $S_{H}^{\rm hard}$, containing only the hard part of the loops, can be obtained from the representation~(\ref{eq:L1loop_final}) by 
expanding the integrand in the hard loop-momentum limit, $p\sim m_H \gg m_L,\,\partial_x$. 
In order to identify the relevant terms 
in this expansion, it is useful to introduce the counting
\begin{align} \label{eq:powerex}
p_\mu,\, m_H \sim\zeta \,,
\end{align}
and determine the order $\zeta^{-k}$, $k>0$, of each term in the integrand of Eq.~(\ref{eq:L1loop_final}). For a given order in $\zeta$ only 
a finite number of terms in the expansion contributes because
$U$ is at most ${\mathcal O}(\zeta)$ and the denominator 
is ${\mathcal O}(\zeta^2)$.\footnote{The part of the operator $U$ coming from $\Delta_H$ arises from interaction terms with at least
three fields. If all three fields are bosons, the dimension-4 operator may contain a dimensionful parameter $\sim \zeta$ or a derivative,
giving rise to a term in $U$ of
${\cal O}(\zeta)$. If two of the fields are fermions the operator is already of dimension 4 and then $\Sigma\sim \zeta^0$, 
which yields a contribution in $U$ of ${\cal O}(\zeta)$ upon application of Eq.~\eqref{eq:Uferm}. Contributions from 
$X^\dagger_{LH}\Delta_L^{-1}X_{LH}$, in the following referred as heavy-light, appear from the product of two interaction 
terms and a light-field propagator and hence they generate terms in $U$ of ${\cal O}(\zeta^0)$.}
For instance, to obtain the dimension-six effective operators, {\em i.e.} those suppressed by $1/m_H^2$, it is enough to truncate the expansion up to terms of $\mathcal{O}\left(\zeta^{-2}\right)$, which means computing $U$ up to $\mathcal{O}\left(\zeta^{-4}\right)$ (recall that $d^4p\sim \zeta^4)$.
Though it was phrased differently, this prescription is effectively equivalent to the one used in
Refs.~\cite{Dittmaier:1995cr,Dittmaier:1995ee} to obtain the non-decoupling effects ({\em i.e.} the ${\cal O}(m_H^0)$ terms)
introduced by a SM-like heavy Higgs.

Finally we recall that, although the covariance of the expansion in Eq.~\eqref{eq:L1loop_final} is not manifest, the symmetry of the functional trace guarantees that the final result can be rearranged such that all the covariant derivatives appear in commutators~\cite{Chan:1986jq,Cheyette:1987qz}. 
As a result, one can always rearrange the expansion of Eq.~\eqref{eq:L1loop_final} in a manifestly covariant way in terms of traces containing powers 
of $U$, field-strength tensors and covariant derivatives acting on them. As noted in Refs.~\cite{Gaillard:1985uh,Cheyette:1987qz,Ball:1988xg},
this rearrangement can be easily performed when $U$ does not
depend on derivatives, as it is the case when only heavy particles enter in the loop\footnote{With the exception of theories with massive vector fields and derivative couplings among two heavy and one light fields.}. However, for the case where $U=U\left(x,\partial_x+ip\right)$, as it happens in general in theories with heavy-light loops, the situation is more involved and the techniques developed in Refs.~\cite{Gaillard:1985uh,Cheyette:1987qz,Ball:1988xg} cannot be directly applied. In this more general case it is convenient to separate $U$ into momentum-dependent and momentum-independent pieces, {\em i.e.} $U=U_H(x)+U_{LH}\left(x,\partial_x+ip\right)$ which, at the diagrammatic level, corresponds to a separation into pure heavy loops and heavy-light loops.
This separation presents two major advantages: first, the power counting for $U_H$ and $U_{LH}$ is generically different, with $U_H$ at most $\mathcal{O}\left(\zeta\right)$ and $U_{LH}$ at most $\mathcal{O}\left(\zeta^0\right)$, both for bosons and fermions, which allows for a different truncation of the series in Eq.~\eqref{eq:L1loop_final} for the terms involving only pure heavy contributions and those involving at least one power of $U_{LH}$. Second, universal expansions of Eq.~\eqref{eq:L1loop_final} in a manifestly covariant form for $U=U_H(x)$ have been derived in the literature up to $\mathcal{O}\left(\zeta^{-2}\right)$, {\em i.e.} for the case of dimension-six operators~\cite{Ball:1988xg,Bilenky:1993bt,Henning:2014wua}, that we reproduce in Eq.~\eqref{eq:Huniversal}. The evaluation of the remaining piece, corresponding to terms containing at least one power of $U_{LH}$ can be done explicitly from Eq.~\eqref{eq:S_Hhard}. 

Let us end the section by summarizing the steps required to obtain the one-loop matching coefficients in our method:
\begin{enumerate}
 \item We collect all field degrees of freedom in $\cal L$, light and heavy, in a field multiplet $\eta=(\eta_H,\eta_L)$, where $\eta_i$ and $(\eta_i)^*$ must be written as separate components for charged fields. We split the fields into classical and quantum part, i.e $\eta \to \hat{\eta} + \eta$, and identify the fluctuation operator $\mathcal{O}$ from the second order variation of
 $\cal L$ with respect to $\eta^*$ and $\eta$ evaluated at the classical field configuration, see Eqs.~\eqref{eq:2ndorderL} and~\eqref{eq:fluctuation_op},
 \begin{align}
 \mathcal{O}_{ij}=\frac{\partial^2\mathcal{L}}{\partial\eta^*_i\,\partial\eta_j}\bigg|_{\eta=\hat\eta}\,.
 \end{align}

 \item We then consider $U(x,\partial_x)$, given in Eqs.~\eqref{eq:Udefined} and~\eqref{eq:Uferm},  with $\widetilde{\Delta}_H$ defined in Eq.~\eqref{eq:tildeDeltaH} in terms of the components of $\mathcal{O}$. Derivatives in $U$ must be shifted as 
 $\partial_x\to\partial_x+ip$. The computation of $U$ requires the inversion of $\Delta_L$: A general expression for the latter is provided in Eq.~\eqref{eq:DeltaL}. The operator $U(x,\partial_x+ip)$ has to be expanded up to a given order in $\zeta$, with the counting given by $p,m_H\sim\zeta \gg m_L, \partial_x$. For deriving the dimension-six EFT operators, the expansion of $U$ must be taken up to $\mathcal{O}\left(\zeta^{-4}\right)$.
 
 \item The final step consists on the evaluation of the traces of $U(x,\partial_x+ip)$ in Eq.~\eqref{eq:L1loop_final} up to the desired order  -- $\mathcal{O}\left(\zeta^{-2}\right)$ for the computation of the one-loop dimension-six effective Lagrangian --. For this computation it is convenient to make the separation $U(x,\partial_x+ip)=U_H(x)+U_{LH}\left(x,\partial_x+ip\right)$ and apply the standard formulas for the traces of $U_H(x)$, see Eq.~\eqref{eq:Huniversal}. The remaining contributions consist in terms involving at least one power of $U_{LH}\left(x,\partial_x+ip\right)$: A general formula for the case of dimension-six operators can be found in Eq.~\eqref{eq:LHuniversal}. Their computation only requires trivial integrals of the form:
\begin{align}\label{eq:master_int}
\begin{aligned}
\int \frac{d^d p}{\left(2\pi\right)^d}\frac{p_{\mu_1}\dots p_{\mu_{2k}}}{(p^{2})^\alpha\left(p^2-m_H^2\right)^\beta}&=\frac{\left(-1\right)^{\alpha+\beta+k}i}{\left(4\pi\right)^{\frac{d}{2}}}\,\frac{\Gamma\left(\frac{d}{2}+k-\alpha\right)\Gamma\left(-\frac{d}{2}-k+\alpha+\beta\right)}{ \Gamma(\beta)\,\Gamma\left(\frac{d}{2}+k\right)}\\[5pt]
&\quad\times\, \frac{g_{\mu_1\dots\mu_{2k}}}{2^k}\, m_H^{d+2k-2\alpha-2\beta}\,,
\end{aligned}
\end{align}
where $g_{\mu_1\dots\mu_{2k}}$ is the totally symmetric tensor with $2k$ indices constructed from $g_{\mu\nu}$ tensors.
 
Terms containing open covariant derivatives, {\em i.e.} derivatives acting only at the rightmost of the traces, should be kept throughout the computation and will either vanish or combine in commutators, yielding gauge-invariant terms with field strength tensors. A discussion about such terms can be found in Appendix~\ref{ap:D6formulae}.
\end{enumerate}

\section{Comparison with previous approaches}\label{s:murayama}

In Ref.~\cite{Henning:2016lyp}, a procedure to obtain the one-loop matching coefficients also using functional integration has been proposed. We wish to highlight here the 
differences of that method, in the following referred as HLM, with respect to the one presented in this manuscript. 

The first difference is how Ref.~\cite{Henning:2016lyp} disentangles contributions from
heavy-light loops from the rest. In the HLM method the determinant of the fluctuation operator ${\cal O}$ which defines
the complete one-loop action $S$ is split using an identity (see their Appendix B) that is formally equivalent 
in our language to performing a field transformation of the form
\begin{align}\label{eq:PHLM}
P_{\mbox{\tiny HLM}}=
\begin{pmatrix}
I & -\Delta_H^{-1} X_{LH}^\dagger \\
0 & I
\end{pmatrix}\,,
\end{align}
that block-diagonalizes the fluctuation operator as: 
\begin{align}\label{eq:field_redef_HLM}
P_{\mbox{\tiny HLM}}^\dagger \, \mathcal{O}P_{\mbox{\tiny HLM}}=
\begin{pmatrix}
\Delta_H & 0\\
0   & \widetilde{\Delta}_L
\end{pmatrix}\,,
\end{align}
where now 
\begin{align}\label{eq:tildeDeltaL}
\widetilde{\Delta}_L = \Delta_L-X_{LH}\Delta_H^{-1}X_{LH}^\dagger\,.
\end{align}
The functional determinant is then separated in the HLM framework into two terms: The determinant of
$\Delta_H$, that corresponds to the loops with only heavy particles, and the determinant of 
$\widetilde{\Delta}_L$, containing both the loops with only light propagators and those with 
mixed heavy and light propagators. The former contributes directly to $U_H$, and provides part of the one-loop
matching conditions (namely those denoted as ``heavy'' in Ref.~\cite{Henning:2016lyp}), upon 
using the universal formula valid for $U$ not depending in derivatives,
Eq.~\eqref{eq:Huniversal}, up to a given order in the expansion in $1/m_H$. 
On the other hand, to obtain the matching conditions that arise from
$\widetilde{\Delta}_L$
(called ``mixed'' contributions in the HLM terminology), 
one has to subtract those contributions already contained in the one-loop terms from the EFT theory 
matched at tree-level. To perform that subtraction without computing both the determinant
of $\widetilde{\Delta}_L$ and that of the quadratic fluctuation of ${\cal L}_{\rm EFT}^{\rm tree}$, HLM 
argues that one has to subtract to the heavy propagators that appear in the computation of $\det \widetilde{\Delta}_L$
the expansion of the heavy propagator to a given order in the limit $m_H\to \infty$. According to HLM,
the subtracted piece builds up the terms (``local counterparts'') that match the loops from   
${\cal L}_{\rm EFT}^{\rm tree}$. These ``local 
counterparts'' have to be identified for each order in the EFT, 
and then dropped prior to the  evaluation of the functional traces. 
This prescription resembles the one used in 
Ref.~\cite{Bilenky:1993bt} to obtain the one-loop effective Lagrangian from integrating out a heavy scalar singlet 
added to the SM. 

While we do not doubt the validity of the HLM method, which the authors of Ref.~\cite{Henning:2016lyp} have shown 
through specific examples,
we believe the framework presented in this manuscript brings some important simplifications. 
Let us note first that in the method of
Ref.~\cite{Henning:2016lyp}, contributions from heavy-light loops are incorporated into 
$\det \widetilde{\Delta}_L$, which results from the functional integration over the 
light fields.
If the light sector contains both bosonic and fermonic degrees of freedom that interact with the heavy sector (as it is the case in most extensions of the SM), a further diagonalization of  
$\widetilde{\Delta}_L$ into bosonic and fermionic blocks is required in order to perform
the Gaussian integral over the light fields. That step is avoided in our approach,
where we shift all heavy particle effects into $\widetilde{\Delta}_H$ and we only need to perform
the path integral over the heavy fields.
Secondly, our method provides a closed
formula (up to trivial integrations which depend on the structure of $U_{LH}$) valid for any
given model, from which the matching conditions of
all EFT operators of a given dimension are obtained. In this sense it is more systematic 
than the 
subtraction prescription of the HLM method, which requires some prior identification of the subtraction terms for the heavy particle 
propagators in the model of interest. Furthermore, in the HLM procedure the light particle mass in the light field propagators is not expanded out in the computation of the functional 
traces, and intermediate results are therefore more involved. In particular, non-analytic terms
in the light masses can appear in intermediate steps of the calculation, and cancellations of
such terms between different contributions have to occur to get the infrared-finite matching coefficients at one loop. Given the amount of algebra involved in the computation of the functional traces, automation
is a prerequisite for integrating out heavy particles in any realistic model. In our method, such automation
is straightforward (and indeed has been used for the heavy real scalar triplet example given in Section~\ref{s:examples}). 
From the description of Ref.~\cite{Henning:2016lyp}, it seems to us that is  harder to implement
the HLM method into an automated code that does not require some manual intervention.

An alternative framework  to obtain the one-loop effective Lagrangian through functional integration, that shares many similarities with that of HLM, has been suggested in 
Ref.~\cite{Ellis:2016enq}. The authors of Ref.~\cite{Ellis:2016enq} have also introduced a subtraction procedure that involves the truncation of the heavy particle propagator. 
Their result for the dimension-6 effective Lagrangian in the  case that  the heavy-light quadratic fluctuation is derivative-independent has been written in terms of traces of
manifestly gauge-invariant operators depending on the quadratic fluctuation $U(x)$, times
coefficients where the EFT contributions have been subtracted. Examples on the calculation
of such subtracted coefficients, which depend on the ultraviolet model, are provided in this reference. 
The approach is however 
limited, as stated by the authors, by the fact that it cannot be applied to cases where the heavy-light interactions contain derivative terms. That is the case, for instance, in extensions of the SM where the heavy fields have interactions with the SM gauge bosons (see the example we provide in Subsection~\ref{ss:example2}). Let us also note that the general formula provided  in the framework of Ref.~\cite{Ellis:2016enq} is written in terms of the components of 
the original fluctuation operator where no diagonalization to separate heavy- and light-field blocks has been performed. This implies that its application to models with mixed statistics in the part of the light sector that interacts with the heavy one, and even  to models where the heavy and light degrees of freedom have different statistics,  must require additional steps that are not discussed in Ref.~\cite{Ellis:2016enq}.

\section{Examples} \label{s:examples}
In this section we perform two practical applications of the framework that we have developed above.
The first one is a scalar toy model simple enough to allow a comparison of our method with
the standard matching procedure.
Through this example we can also illustrate explicitly that matching coefficients arise
from the hard region of the one-loop amplitudes in the full theory. The second example corresponds
to a more realistic case where one integrates out a heavy real scalar triplet that has been
added to the SM. 
\subsection{Scalar toy model} \label{ss:example1}
Let us consider a model with two real scalar fields, $\varphi$ with mass $m$ and $\phi$ with mass $M$, whose interactions are described by the Lagrangian
\begin{equation} \label{eq:tphi4}
{\cal L}(\varphi,\phi) = \frac{1}{2} \left( \partial_{\mu}\phi \,  \partial^{\mu} \phi - M^2 \,  \phi^2 \right) +
 \frac{1}{2} \left( \partial_{\mu}\varphi \, \partial^{\mu} \varphi - m^2 \, \varphi^2 \right)  - \frac{\kappa}{4!} \, \varphi^4 - \frac{\lambda}{3 !} \, \varphi^3  \, \phi \, .
\end{equation} 
Assuming $M \gg m$ we wish to determine the effective field theory resulting from integrating out the $\phi$ field: ${\cal L}_{\mbox{\tiny{EFT}}}(\hat\varphi)$. We
perform the calculation up to and including $1/M^{2}
$-suppressed operators in the EFT. Within this model this implies that we have to consider up to six-point Green functions. 
This same model has also been considered in Ref.~\cite{Henning:2016lyp}.
\par 
At tree level we solve for the equation of motion of the $\phi$ field and we obtain
\begin{align}
\hat{\phi} = - \frac{\lambda}{6M^2} \, \hat{\varphi}^3 + {\cal O}(M^{-4})\,,
\end{align}
that, upon substituting in Eq.~\eqref{eq:tphi4}, gives the tree-level effective Lagrangian
\begin{equation} \label{eq:ttree}
{\cal L}^{\rm tree}_{\mbox{\tiny{EFT}}} =  \frac{1}{2} \left( \partial_{\mu}\hat \varphi \, \partial^{\mu} \hat \varphi - m^2 \, \hat \varphi^2 \right)  - \frac{\kappa}{4!} \, \hat \varphi^4 +\frac{\lambda^2}{72 M^2} \, \hat{\varphi}^6 \, . 
\end{equation}
To proceed at one loop we use the background field method as explained in Section~\ref{s:method}: $\phi \rightarrow \hat{\phi} + \phi$ and $\varphi \rightarrow \hat{\varphi} + \varphi$. We have $\eta = \left( \phi, \varphi \right)^\intercal$ and we consider the same counting as in Eq.~(\ref{eq:powerex}): $p_{\mu}, M \sim \zeta$. The fluctuation operator in Eq.~(\ref{eq:fluctuation_op}) is given by
\begin{align} \label{eq:tfluctua}
\begin{aligned}
\Delta_H &= - \partial^2 - M^2 \, , \\
\Delta_L &= - \partial^2  - m^2 - \frac{\kappa}{2} \,  \hat{\varphi}^2 - \lambda \,  \hat{\varphi} \, \hat{\phi} \, , \\
X_{LH} &= - \frac{\lambda}{2} \, \hat{\varphi}^2 \, , 
\end{aligned}
\end{align}  
that only depends on the classical field configurations. In order to construct $\widetilde{\Delta}_H(x,\partial_x+ip)$ in Eq.~(\ref{eq:tildeDeltaH}) we need to determine $\Delta_L^{-1}(x,\partial_x+ip)$ up
to, and including, terms of order  $\zeta^{-4}$:
\begin{align} \label{eq:tdeltal}
\begin{aligned}
\Delta_L(x,\partial_x+ip) &= p^2-m^2 - 2 i \,  p_{\mu} \,  \partial^{\mu} - \partial^2 - \frac{\kappa}{2} \, \hat{\varphi}^2 - \lambda \, \hat{\varphi}
\, \hat{\phi} \, ,\\  
\Delta_L^{-1} (x,\partial_x + ip) &= \frac{1}{p^2} \left( 1 + \frac{m^2}{p^2} \right) +  \frac{1}{p^4} \left( 2 i \, p_{\mu}  \partial^{\mu} + 
\partial^2 + \frac{\kappa}{2} \hat{\varphi}^2 \right) - 4 \, \frac{p_{\mu} p_{\nu}}{p^6} \,  \partial^{\mu} \partial^{\nu} + {\cal O} (\zeta^{-5} ) \, .
\end{aligned}
\end{align}
Using this result we get $U(x,\partial_x+ip)$ from Eq.~(\ref{eq:Udefined})
\begin{equation} \label{eq:tU}
U(x,\partial_x+ip) = \frac{\lambda^2}{4} \, \hat{\varphi}^2 \, \left[ \frac{1}{p^2} \left( 1 + \frac{m^2}{p^2} \right) +  \frac{1}{p^4} \left( 2 i \, p_{\mu}  \partial^{\mu} + 
\partial^2 + \frac{\kappa}{2} \hat{\varphi}^2 \right) - 4 \, \frac{p_{\mu} p_{\nu}}{p^6} \,  \partial^{\mu} \partial^{\nu} \right] \, \hat{\varphi}^2 + {\cal O} (\zeta^{-5} )\, . 
\end{equation}
Inserting this operator in Eq.~(\ref{eq:L1loop_final}), we notice that at the order we are considering only the $n=1$ term contributes,  with
\begin{equation} \label{eq:tleff}
{\cal L}_{\mbox{\tiny{EFT}}}^{\rm 1loop} \, = \, - \frac{i}{2}  \int \frac{d^dp}{(2 \pi)^d} \, \frac{U(x,\partial_x+ip)}{p^2-M^2} \, .
\end{equation}
The momentum integration can be readily performed: In the $\overline{\rm MS}$ regularization scheme with $\mu=M$ we finally obtain
\begin{equation} \label{eq:t46result}
{\cal L}_{\mbox{\tiny{EFT}}}^{\rm 1loop} \, = \, \frac{\lambda^2}{16 (16 \pi^2)} \left[ 2 \left( 1 + \frac{m^2}{M^2} \right) \hat{\varphi}^4
- \frac{1}{M^2} \hat{\varphi}^2 \partial^2 \hat{\varphi}^2 + \frac{\kappa}{M^2} \hat{\varphi}^6 \right] \, . 
\end{equation} 

Let us recover now this result through the usual matching procedure between the full theory ${\cal L}(\varphi,\phi)$ in Eq.~(\ref{eq:tphi4}) and the
effective theory without the heavy scalar field $\phi$. 
Our goal is to further clarify the
discussion given in Section~\ref{s:method} on the hard origin of the matching 
coefficients of the effective theory by considering this purely academic case.
In order to make contact with the result obtained in Eq.~(\ref{eq:t46result}) using the
functional approach, we perform the matching off-shell and we use the $\overline{\rm MS}$ regularization scheme with $\mu=M$. 
We do not consider in the matching procedure one-loop diagrams with only light fields, since they are present in both
the full-theory and the effective theory amplitudes and, accordingly, cancel out in the matching.
\par 
For the model under discussion there is no contribution to the two- and three-point Green functions involving heavy particles in the loop. The diagrams contributing to the matching of the four-point Green function are given by
\begin{fmffile}{toy_model_4PUV}
\begin{align}\label{eq:4PUV}
\begin{aligned}
\begin{gathered}
\begin{fmfgraph*}(80,45)\fmfkeep{d1_4PUV}
\fmfleft{i1,i2}
\fmfright{o1,o2}
\fmf{plain}{i1,v1}
\fmf{plain}{i2,v1}
\fmf{dashes,left=1.0,tension=0.65}{v1,v2}
\fmf{plain,right=1.0,tension=0.65}{v1,v2}
\fmf{plain}{v2,o1}
\fmf{plain}{v2,o2}
\end{fmfgraph*}
\end{gathered}&=\frac{i}{16 \pi^2} \,  \lambda^2 \left[ 3 + 3 \, \frac{m^2}{M^2} + \frac{s+t+u}{2 M^2} \right]\Bigg|_{\rm hard} \,\\
&\quad+\frac{i}{16 \pi^2} \, \lambda^2 \left[ - 3 \,  \frac{m^2}{M^2} + 3 \, \frac{m^2}{M^2} \ln \left( \frac{m^2}{M^2} \right) \right]\Bigg|_{\rm soft} + {\cal O}(M^{-4}) ,  \\[10pt]
\begin{gathered}
\begin{fmfgraph*}(80,45)\fmfkeep{d2_4PUV}
\fmfleft{i1,i2,i3}
\fmfright{o1,op}
\fmf{plain,tension=0.4}{i1,v1}
\fmf{plain,tension=0.4}{i2,v1}
\fmf{plain,tension=0.4}{i3,v1}
\fmf{dashes}{v1,v2}
\fmf{plain,tension=0.7,right}{v2,v2}
\fmf{plain}{v2,o1}
\fmf{phantom,left}{v2,op}
\end{fmfgraph*}
\end{gathered}&=\frac{i}{16 \pi^2} \, \lambda^2 \left[ - 2  \, \frac{m^2}{M^2} + 2 \, \frac{m^2}{M^2} \ln \left( \frac{m^2}{M^2} \right) \right]\Bigg|_{\rm soft} + {\cal O}(M^{-4}) ,
\end{aligned}
\end{align}
\end{fmffile}

\noindent
where we have explicitly separated the contributions from the hard and soft loop-momentum regions. Note that a non-analytic term in $m$ can
only arise from the soft region, since in the hard region the light mass and the external momenta are expanded out from the propagators. 
For the corresponding EFT computation we need the effective Lagrangian matched at one-loop:
\begin{align}\label{eq:EFT_lag}
{\cal L}_{\mbox{\tiny{EFT}}}={\cal L}^{\rm tree}_{\mbox{\tiny{EFT}}}
 + \frac{\alpha}{4!} \, \hat\varphi^4 \, + \, \frac{\beta}{4! M^2} \, \hat\varphi^2 \partial^2 \hat\varphi^2
 + \frac{\gamma}{6! M^2} \, \hat\varphi^6
\,,
\end{align}
which now includes the dimension-6 operator with four light 
fields, and the one-loop matching coefficient for the  
4- and 6-light field operators already present in ${\cal L}^{\rm tree}_{\mbox{\tiny{EFT}}}$.
The EFT contributions to the four-point Green function read
\begin{fmffile}{toy_model_4PEFT}
\begin{align}\label{eq:4PEFT}
\begin{aligned}
\begin{gathered}
\begin{fmfgraph*}(80,45)\fmfkeep{d1_4PEFT}
\fmfleft{i1,i2}
\fmfright{o1,o2}
\fmf{plain}{i1,v1}
\fmf{plain}{i2,v1}
\fmf{plain,tension=0.9,up}{v1,v1}
\fmf{plain}{v1,o1}
\fmf{plain}{v1,o2}
\fmfv{decor.shape=circle,decor.filled=full,decor.size=2.5mm}{v1}
\end{fmfgraph*}
\end{gathered}&=\frac{i}{16 \pi^2} \, \lambda^2 \left[ -5 \, \frac{m^2}{M^2} + 5 \, \frac{m^2}{M^2} \log \left( \frac{m^2}{M^2} \right) \right] \, 
+ {\cal O}(M^{-4}) \, , \\[10pt]
\begin{gathered}
\begin{fmfgraph*}(80,45)\fmfkeep{d2_4PEFT}
\fmfleft{i1,i2}
\fmfright{o1,o2}
\fmf{plain}{i1,v1}
\fmf{plain}{i2,v1}
\fmf{plain}{v1,o1}
\fmf{plain}{v1,o2}
\fmfv{decor.shape=square,decor.filled=full,decor.size=2.5mm}{v1}
\end{fmfgraph*}
\end{gathered}&=i \, \alpha \, - \, i \,  \frac{\beta}{3 M^2} \, \left( s+t+u \right) \,   .
\end{aligned}
\end{align}
\end{fmffile}

\noindent
We see that the soft components of the full-theory amplitude match the one-loop diagram in the effective theory, 
and the matching coefficients of the $\varphi^4$ operators get thus determined by the hard part of the one-loop full-theory amplitude:
\begin{eqnarray} \label{eq:tresalbet}
\alpha \, = \, \frac{3}{16 \pi^2} \, \lambda^2  \left( 1 + \frac{m^2}{M^2} \right) \;   & , &  \; \; \;  \; \; \;  \beta \, = \,  - \, \frac{3}{16 \pi^2} \, \frac{\lambda^2}{2} \,  . 
\end{eqnarray}
in agreement with the result for the $\varphi^4$ terms in Eq.~(\ref{eq:t46result}).
\par 
The next contribution to the one-loop effective theory comes from the six-point Green function. The full theory provides two diagrams for the matching:
\begin{fmffile}{toy_model_6PUV}
\begin{align}\label{eq:6PUV}
\begin{aligned}
\begin{gathered}
\begin{fmfgraph*}(80,45)\fmfkeep{d1_6PUV}
\fmfleft{i1,i2,i3}
\fmfright{o1,o2,o3}
\fmf{plain,tension=2}{i2,v1}
\fmf{plain,tension=2}{i3,v1}
\fmf{dashes,tension=2}{v1,v2}
\fmf{plain,tension=0.5}{v1,v3}
\fmf{plain,tension=0.5}{v3,v2}
\fmf{plain,tension=2}{v3,i1}
\fmf{plain,tension=2}{v3,o1}
\fmf{plain,tension=2}{v2,o2}
\fmf{plain,tension=2}{v2,o3}
\end{fmfgraph*}
\end{gathered}&\,=\frac{i}{16 \pi^2} \,  45 \, \frac{\kappa \, \lambda^2}{M^2}\Bigg|_{\rm hard}+\frac{i}{16 \pi^2} \,  45 \, \frac{\kappa \, \lambda^2}{M^2} \ln \left( \frac{m^2}{M^2} \right) \Bigg|_{\rm soft} + {\cal O}(M^{-4}) , \\[10pt]
\begin{gathered}
\begin{fmfgraph*}(80,45)\fmfkeep{d2_6PUV}
\fmfleft{i1,i2,i3}
\fmfright{o1,op1,o2,o3}
\fmf{plain,tension=0.4}{i1,v1}
\fmf{plain,tension=0.4}{i2,v1}
\fmf{plain,tension=0.4}{i3,v1}
\fmf{dashes,tension=0.65}{v1,v2}
\fmf{plain,tension=0.25,right}{v2,v3}
\fmf{plain,tension=0.25,right}{v3,v2}
\fmf{plain,tension=0.3}{v3,o2}
\fmf{plain,tension=0.3}{v3,o3}
\fmf{plain,tension=0.05}{v2,o1}
\fmf{phantom,tension=0.4}{v2,op1}
\end{fmfgraph*}
\end{gathered}&=\frac{i}{16 \pi^2} \, 30 \, \frac{\kappa \, \lambda^{2}}{M^2} \ln \left( \frac{m^2}{M^2} \right)\Bigg|_{\rm soft} + {\cal O}(M^{-4}) ,
\end{aligned}
\end{align}
\end{fmffile}

\noindent
where once more we have explicitly separated the hard and soft contributions from each diagram. The six-point effective theory amplitude gives
\begin{fmffile}{toy_model_6PEFT}
\begin{align}\label{eq:6PEFT}
\begin{aligned}
\begin{gathered}
\begin{fmfgraph*}(80,45)\fmfkeep{d1_6PEFT}
\fmfleft{i1,i2,i3}
\fmfright{o1,o2,o3}
\fmf{plain,tension=4}{i1,v1}
\fmf{plain,tension=4}{i2,v1}
\fmf{plain,tension=0.5,right}{v1,v2}
\fmf{plain,tension=0.5,right}{v2,v1}
\fmf{plain,tension=4}{v2,i3}
\fmf{plain,tension=4}{v2,o3}
\fmf{plain,tension=4}{v1,o1}
\fmf{plain,tension=4}{v1,o2}
\fmfv{decor.shape=circle,decor.filled=full,decor.size=2.5mm}{v1}
\end{fmfgraph*}
\end{gathered}&=\frac{i}{16 \pi^2} \,  75 \, \frac{\kappa \, \lambda^2}{M^2} \ln \left( \frac{m^2}{M^2} \right) \, 
+ {\cal O}(M^{-4}) \, , \\[10pt]
\begin{gathered}
\begin{fmfgraph*}(80,45)\fmfkeep{d2_6PEFT}
\fmfleft{i1,i2,i3}
\fmfright{o1,o2,o3}
\fmf{plain}{i1,v1}
\fmf{plain}{i2,v1}
\fmf{plain}{i3,v1}
\fmf{plain}{v1,o1}
\fmf{plain}{v1,o2}
\fmf{plain}{v1,o3}
\fmfv{decor.shape=square,decor.filled=full,decor.size=2.5mm}{v1}
\end{fmfgraph*}
\end{gathered}&=i \frac{\gamma}{M^2} \, .
\end{aligned}
\end{align}
\end{fmffile}

\noindent
Again, we note that the soft terms of the full theory are reproduced by the one-loop diagram in the effective theory. The local contribution is determined by the hard part of the full theory amplitude and thus reads
\begin{equation} \label{eq:tresgamma}
\gamma \, = \, \frac{45}{16 \pi^2} \, \kappa \, \lambda^2 \, ,
\end{equation}
that matches the result found in Eq.~(\ref{eq:t46result}) for the $\hat\varphi^6$ term. 

\subsection{Heavy real scalar triplet extension}\label{ss:example2}
As a second example, we consider 
an extension of the SM with an extra scalar sector comprised by a triplet of heavy scalars with zero hypercharge, $\Phi^a, a=1,2,3$, 
which interacts with the light Higgs doublet
\cite{Khandker:2012zu}. A triplet of scalars are ubiquitous in many extensions of the SM since the seminal article by
Gelmini and Roncadelli \cite{Gelmini:1980re}. However, we are not interested here in the phenomenology of the model but
in how to implement our procedure in order to integrate out, at one loop, the extra scalar sector of the theory, assumed it is
much heavier than the rest of the spectrum.
Partial results for the dimension-6 operators involving the
light Higgs doublet that are generated from this model
have been provided in the functional approaches
of Refs.~\cite{Henning:2016lyp,Ellis:2016enq}.
\par 
The Lagrangian of the model is given by
\begin{align} \label{eq:triplets}
\mathcal{L}=\mathcal{L}_{\mbox{\scriptsize SM}}+ \frac{1}{2}D_\mu\Phi^aD^\mu\Phi^a-\frac{1}{2}M^2\Phi^a\Phi^a-\frac{\lambda_\Phi}{4}\left(\Phi^a\Phi^a\right)^2 
+\kappa\,\left(\phi^\dagger\tau^a\phi\right)\Phi^a-\eta\left(\phi^\dagger\phi\right) \Phi^a\Phi^a\,,
\end{align}
Here $\phi$ is the SM Higgs doublet and the covariant derivative acting on the triplet is defined as $D_{\mu} \Phi^a \equiv D_\mu^{ac}\, \Phi^c = \left(\partial_{\mu}\,\delta^{ac} + g \varepsilon^{abc} W_{\mu}^b\right) \Phi^c$.  Within the background field method we split the fields into their classical (with hat) and quantum components: $\Phi^a \rightarrow  \hat\Phi^a + \Phi^a$, $\phi \rightarrow  \hat\phi +\phi$ and $W_{\mu}^a \rightarrow \hat{W}_{\mu}^a + W_{\mu}^a$. Given as an expansion in inverse powers of its mass, the classical field of the scalar triplet reads
\begin{align} \label{eq:eomphi}
\begin{aligned}
\hat\Phi^a=\,\frac{\kappa}{M^2}\left(\hat\phi^\dagger\tau^a\hat\phi\right)-\frac{\kappa}{M^4}\left[\hat D^2+2\eta\left(\hat\phi^\dagger\hat\phi\right)\right]\left(\hat\phi^\dagger\tau^a\hat\phi\right)
+\mathcal{O}\left(\frac{\kappa}{M^6}\right)\,.
\end{aligned}
\end{align}
Following the procedure described in the Section~\ref{s:method} we divide the fields into heavy and 
light, respectively, as $\eta_H = \Phi^a$ and $\eta_L = \{ \phi, \phi^*, W_{\mu}^a \}$. The fluctuation matrix is
readily obtained from Eqs.~\eqref{eq:2ndorderL} and~\eqref{eq:fluctuation_op},
\begin{align} \label{eq:fluctuatri}
\begin{aligned}
\Delta_H&=\Delta_{\Phi\Phi}^{ab}\,,\\
X_{LH}^{\dagger}&=
\left(
\left(X_{\phi^*\Phi}^{a}\right)^{\dagger} \;  \left(X_{\phi^*\Phi}^{a}\right)^\intercal \;  \big(X_{W\Phi}^{\nu\, da}\big)^\intercal
\right),\\
\Delta_{L}&=
\begin{pmatrix}
\Delta_{\phi^*\phi} & X_{\phi\phi}^\dagger & \left(X_{W\phi}^{\nu\,d}\right)^\dagger\\[5pt]
X_{\phi\phi} & \Delta_{\phi^*\phi}^\intercal & \left(X_{W\phi}^{\nu\,d}\right)^\intercal\\[5pt]
X_{W\phi}^{\mu\,c} & \left(X_{W\phi}^{\mu\,c}\right)^* & \Delta_W^{\mu\nu\,cd}
\end{pmatrix}\,,
\end{aligned}
\end{align}
with
\begin{align} \label{eq:fluctus}
\begin{aligned}
\Delta_W^{\mu\nu\,ab}=&\,\left(\Delta_W^{\mu\nu\,ab}\right)_{\mbox{\scriptsize SM}}+g^2\,g_{\mu\nu}\,\epsilon^{acm}\epsilon^{bdm}\,\hat{\Phi}^c\hat{\Phi}^d,\\
\Delta_{\phi^*\phi}=&\left(\Delta_{\phi^*\phi}\right)_{\mbox{\scriptsize SM}}+\kappa\,\tau^a\hat{\Phi}^a-\eta\,\hat{\Phi}^a\hat{\Phi}^a,\\
\Delta_{\Phi\Phi}^{ab}=&\,-\hat{D}_{ab}^2+\delta_{ab}\left[-M^2-\lambda_\Phi\hat{\Phi}^c\hat{\Phi}^c-2\eta\left(\hat{\phi}^\dagger\hat{\phi}\right)\right] -2\lambda_\Phi\hat{\Phi}^a\hat{\Phi}^b\,,\\
X_{W\Phi}^{\mu\, ab}=&\,g\epsilon^{abc}\left(\hat{D}^\mu\hat{\Phi}^c\right)+g\epsilon^{acd}\,\hat\Phi^c\hat{D}^{\mu\,db},\\
X_{\phi^*\Phi}^{a}=&\,\kappa\,\tau^a\,\hat\phi-2\eta\,\hat\phi\,\hat\Phi^a,
\end{aligned}
\end{align}
and the rest of fluctuations in $\Delta_L$ involving only the light fields are contained
in the quadratic piece of the SM Lagrangian, which we provide in 
Eqs.~\eqref{eq:SM_1loop} and~\eqref{eq:SM_fluctuations}.
The quadratic term containing all fluctuations related to the heavy triplet is given
by our formula~\eqref{eq:tildeDeltaH},
\begin{align}\label{eq:Tdet_exact}
\widetilde{\Delta}_{\Phi\Phi}=\Delta_{\Phi\Phi}-X_{LH}^\dagger\Delta_{L}^{-1}X_{LH}\,.
\end{align}
The expansion in inverse powers of the heavy mass of the triplet requires a counting analogous to the one in Eq.~(\ref{eq:powerex}), {\em i.e.} $p_{\mu} \sim \zeta$ and
$M \sim \zeta$. For the counting of the dimensionful parameter $\kappa$ we choose 
$\kappa \sim \zeta$ and then, from
Eq.~(\ref{eq:eomphi}) we have $\hat\Phi^a \sim \zeta^{-1}$.
As we are interested in dimension-six effective operators we can neglect contributions 
$\mathcal{O}\left(\zeta^{-5}\right)$ and smaller. This is because
in Eq.~(\ref{eq:L1loop_final}) the propagator in the heavy particle provides an extra power $\zeta^{-2}$. Hence we only need the numerator up to 
${\cal O}(\zeta^{-4})$.
\par 
For practical reasons we choose to work in the Landau gauge for the quantum fluctuations, {\em i.e.} the renormalizable gauge with $\xi_W = 0$ in 
Eqs.~\eqref{eq:ghost_Lag} and~\eqref{eq:SM_inv_prop}. The computation is much simpler in this gauge because the inverse of the propagators are transverse. Rearranging the expression in Eq.~\eqref{eq:Tdet_exact}, we  can write
\begin{align} \label{eq:fluctuatrip}
\begin{aligned}
\widetilde{\Delta}_{\Phi\Phi}^{ab}& = \Delta_{\Phi\Phi}^{ab}-\left[\left(X_{\phi^*\Phi}^{a}\right)^\dagger\overline\Delta_{\phi^*\phi}^{\;-1}X_{\phi^*\Phi}^{b}
+\left(X_{\phi^*\Phi}^{a}\right)^\intercal\overline X_{\phi\phi}X_{\phi^*\Phi}^{b}+c.c.\right] \\
&\quad -\left(\overline X_{W\Phi}^{\mu\, ca}\right)^\intercal\left(\Delta_W^{\mu\nu\,cd}\right)^{-1}\overline X_{W\Phi}^{\nu\, db} \, + \, \mathcal{O}\left(\zeta^{-5}\right)\,,
\end{aligned}
\end{align}
where $c.c.$ is short for complex conjugation and we have used the following definitions:
\begin{align} \label{eq:fltripdef}
\begin{aligned}
\overline\Delta_{\phi^*\phi}^{\;-1}&=\Delta_{\phi^*\phi}^{-1}+\Delta_{\phi^*\phi}^{-1}X_{\phi\phi}^\dagger\left(\Delta_{\phi^*\phi}^{-1}\right)^\intercal X_{\phi\phi}\Delta_{\phi^*\phi}^{-1}\,,\\
\overline X_{\phi\phi}&=-\left(\Delta_{\phi^*\phi}^{-1}\right)^\intercal X_{\phi\phi}\;\Delta_{\phi^*\phi}^{-1}\,,\\
\overline X_{W\Phi}^{\mu\,ab}&=X_{W\Phi}^{\mu\,ab}-\left(X_{W\phi}^{\mu\,a}\Delta_{\phi^*\phi}^{-1}X_{\phi^*\Phi}^{b}+c.c.\right)\,.
\end{aligned}
\end{align}
To proceed we now come back to Eq.~(\ref{eq:L1loop_final}) (with negative sign), with $m_H = M$ and $U = - \hat{D}^2-M^2-\widetilde{\Delta}_{\Phi \Phi}$. Remember that the hat on the covariant derivatives indicates that only the classical field configuration for the gauge bosons is involved. Then by computing Eq.~(\ref{eq:fluctuatrip}) up to ${\cal O}(\zeta^{-4})$ one can obtain the 
one-loop effective theory that derives from the model specified in Eq.~(\ref{eq:triplets}) upon integrating out the triplet of heavy scalars. 
\par 
We do not intend here to provide the complete result of the generated dimension-six operators.
As a simple example and for illustrative purposes, we provide details on the computation of the heavy-light contributions arising from the quantum fluctuations of the
electroweak gauge bosons. The latter provide the matching contributions to the
dimension-six operators with Higgs fields and no field strength tensors proportional
to $g^2$, which were not obtained with the functional approach in Ref.~\cite{Ellis:2016enq} due to the presence of ``open'' covariant derivatives. The computation of such contributions was
also absent in the approach of Ref.~\cite{Henning:2016lyp}. The relevant 
term in $U(x,\partial_x+ip)$ for this calculation is
\begin{align}\label{eq:barXWbarX}
\left[\left(\overline X_{W\Phi}^{\mu\, ca}\right)^\intercal\left(\Delta_W^{\mu\nu\,cd}\right)^{-1}\overline X_{W\Phi}^{\nu\, db}\right]\left(x,\partial_x+ip\right)\,.
\end{align}
The first operator in Eq.~(\ref{eq:barXWbarX}) simply reads
\begin{align}
\overline X_{W\Phi}^{\mu\,ab}\left(x,\partial_x+ip\right)&=-ig\,\epsilon^{abc}\,\hat\Phi^cp^\mu+\frac{\kappa}{p^2}\,ig\,\epsilon^{abc}\left(\hat\phi^\dagger\tau^c\hat\phi\right)p_\mu \nonumber\\
&\quad +g\epsilon^{abc}\left(\hat{D}^\mu\hat{\Phi}^c\right)+g\epsilon^{acd}\,\hat\Phi^c\hat{D}^{\mu\,db}-\frac{1}{p^2}\,g\kappa\left\{-\frac{i}{2}\left[\left(\hat D_\mu\hat{\phi}\right)^\dagger\tau^a\tau^b\,\hat\phi\right]\right. \nonumber\\
&\quad\left.+\frac{i}{2}\left(\hat{\phi}^\dagger\tau^a\tau^b\hat{D}_\mu\hat\phi\right)+\frac{i}{2}\left(\hat\phi^\dagger\tau^a\tau^d\hat\phi\right)\hat D_\mu^{db}+c.c.\right\}+\mathcal{O}\left(\zeta^{-2}\right) \nonumber\\
&=g\,\epsilon^{abc}\left(\hat{D}^\mu\hat{\Phi}^c\right)-\frac{g\kappa}{p^2}\,i\,\delta_{ab}\Big(\hat{\phi}^\dagger\lra{\hat{D}_\mu}\hat\phi\Big)-g\,\frac{p^2-M^2}{p^2}\,\epsilon^{acd}\,\hat{\Phi}^d\left(\hat D_\mu^{cb}+ip_\mu\,\delta^{cb}\right) \nonumber\\
&\quad+\mathcal{O}\left(\zeta^{-2}\right)\,,
\end{align}
where, in the last line, we used the EOM for the heavy triplet, Eq.~\eqref{eq:eomphi}, and we defined the hermitian derivative terms
\begin{align}
\left(\phi^\dagger\lra{D}_\mu\phi\right)\equiv\left(\phi^\dagger D_\mu\phi\right)-\left[\left(D_\mu\phi\right)^\dagger\phi\right]\,,
\end{align}
with the covariant derivative acting on the Higgs field as specified in Eq.~(\ref{eq:SM_cdev}).
The contributions from the heavy triplet to the fluctuation $\Delta_W$, see Eq.~\eqref{eq:fluctus}, do not affect the computation of $\Delta_W^{-1}\left(x,\partial_x+ip\right)$ at leading order, and we can take the expression given in Eq.~\eqref{eq:SM_inv_prop} (with $\xi_W = 0$) for the latter. As a result we obtain
\begin{align}\label{eq:XdWX}
\left[\left(\overline X_{W\Phi}^{\mu\, ca}\right)^\intercal\left(\Delta_W^{\mu\nu\,cd}\right)^{-1}\overline X_{W\Phi}^{\nu\, db}\right]\left(x,\partial_x+ip\right)&=g^2\left[-\frac{g^{\mu\nu}}{p^2}+\frac{p^\mu p^\nu}{p^4}\right]\left[\delta_{ab}\left(\hat{D}_\mu\hat{\Phi}^c\right)\left(\hat{D}_\nu\hat{\Phi}^c\right)\right.\nonumber\\
&\quad\,\left.-\left(\hat{D}_\mu\hat{\Phi}^a\right)\left(\hat{D}_\nu\hat{\Phi}^b\right)-\,\delta_{ab}\frac{\kappa^2}{p^4}\left(\hat{\phi}^\dagger\lra{\hat{D}_\mu}\hat\phi\right)\left(\hat{\phi}^\dagger\lra{\hat{D}_\nu}\hat\phi\right)\right] \nonumber\\
&\quad\,+\mathcal{O}\left(\zeta^{-5}\right)\,,
\end{align}
and we dropped the terms proportional to $(p^2-M^2)$ since they yield a null contribution in the momentum integration, as explained below.

Only the first term of the series in Eq.~(\ref{eq:L1loop_final}) contributes in this case:
\begin{equation} \label{eq:Tgleff}
\left.\cal L_{\mbox{\tiny{EFT}}}^{\rm 1loop}\right|_W
 \, = \, 
 - \frac{i}{2}  \int \frac{d^dp}{(2 \pi)^d} \, \frac{\left[\left(\overline X_{W\Phi}^{\mu\, ca}\right)^\intercal\left(\Delta_W^{\mu\nu\,cd}\right)^{-1}\overline X_{W\Phi}^{\nu\, da}\right]\left(x,\partial_x+ip\right)}{p^2-M^2} \, .
\end{equation}
From Eq.~(\ref{eq:Tgleff}) it is clear that terms proportional to $(p^2-M^2)$
yield scaleless  terms that are set to zero in dimensional regularization, which justifies having dropped them in Eq.~\eqref{eq:XdWX}. After evaluating the integral in the $\overline{\rm MS}$ regularization scheme, 
using the heavy triplet EOMs and rearranging the result through partial integration we finally get for $\mu=M$
\begin{equation} \label{eq:Tgleff_final}
\left.\cal L_{\mbox{\tiny{EFT}}}^{\rm 1loop}\right|_W \, = \, 
\frac{1}{16\pi^2}\frac{g^2\kappa^2}{M^4}\left[-\frac{25}{16}\left(\hat\phi^\dagger\hat\phi\right)\partial^2\left(\hat\phi^\dagger\hat\phi\right)+\frac{5}{4}\left[\left(\hat\phi^\dagger\hat\phi\right)\left(\hat\phi^\dagger \hat D^2\hat\phi\right)+h.c.\right]-\frac{5}{4}\left|\hat\phi^\dagger \hat D_\mu\hat\phi\right|^2\right]\,.
\end{equation}
In order to compare this result with previous calculations done in the literature, we focus on the heavy triplet contributions to $Q_{\phi D} = \left| \phi^{\dagger} D^{\mu} \phi \right|^2$. From the result in Eq.~\eqref{eq:Tgleff_final} we find for its one-loop matching coefficient
\begin{align}
C_{\phi D}(\mu = M)\Big|_{{\cal O}(g^2)} = 
-\frac{1}{16\pi^2}\frac{\kappa^2}{M^4} \frac{5}{4} g^2\,,
\end{align}
which agrees with the result given in Ref.~\cite{delAguila:2016zcb} for the term proportional to $g^2$. 
The remaining contributions to $C_{\phi D}(\mu = M)$ have also been calculated with our method. However their computation is lengthy and does not provide any new insight on the method. The final result reads
\begin{equation} \label{eq:Tresult}
C_{\phi D}(\mu = M) = \frac{\kappa^2}{M^4} \left[ -2 \, + \, \frac{1}{16 \pi^2} \left( 5 \frac{\kappa^2}{M^2} - \frac{5}{4} g^2 + 16 \eta - 3 \lambda
- 20 \lambda_{\Phi} \right) \right] . 
\end{equation}
In Eq.~(\ref{eq:Tresult}) we have also included the term arising from the redefinition of $\phi$ that absorbs the one-loop contribution to the kinetic term, 
$\phi \to \big( 1 -3\kappa^2/64\pi^2M^2 \big) \phi$. This result is in agreement with the one provided in Ref.~\cite{delAguila:2016zcb} once we account for the different convention in the definition of $\lambda$: our $\lambda$ equals $2 \lambda$ in that reference.

\section{Conclusions}\label{s:conclusions}
The search for new physics in the next run at LHC stays as a powerful motivation 
for a systematic scrutiny of the possible extensions of the SM. The present status that
engages both collider and precision physics has, on the theoretical side, a robust tool in the construction, treatment and phenomenology of effective
field theories that are the remains of ultraviolet completions of the SM upon integration of heavy spectra.  
\par 
Though, traditionally, there are two essential procedures to construct those effective field theories,
namely functional methods and matching schemes, the latter have become the most frequently used. Recently there has been a rediscovery of the
functional methods, initiated by the work of Henning et al.~\cite{Henning:2014wua}.
The latter work started  a discussion regarding the treatment of the terms that mix heavy and light quantum fluctuations, that was finally clarified but which, in our opinion, was already settled in the past literature on the subject. 
In this article we have addressed
this issue and we have provided a framework that further clarifies the treatment
of the heavy-light contributions and simplifies the technical modus operandi.
\par 
The procedure amounts to a particular diagonalization of the quadratic form in the path integral 
of the full theory that leaves untouched the part that entails the light 
fields. In this way we can integrate, at one loop, contributions with only heavy fields inside the loop and contributions with mixed components of heavy
and light fields, with a single computation and following the conventional method employed to carry out the first ones only. 
We have also showed that in the resulting determinant 
containing the heavy particle effects only the 
{\em hard} components are needed to derive the 
one-loop matching coefficients of the effective theory. Within dimensional regularization
these hard contributions are obtained by expanding out the low-energy scales with 
respect the hard loop momentum which has to be considered of the same order
as the mass of the heavy particle. In this way, our determination of the EFT local
terms that reproduce the heavy-particle
effects does not require the
subtraction of any one-loop contributions from the EFT, as opposed to 
the conventional (diagrammatic) matching approach or to the recently proposed
methods that use functional techniques.
We have included two examples in Section~\ref{s:examples}: A scalar toy model,
that nicely illustrates the simplicity of our approach as compared to the diagrammatic
approach, and a heavy real scalar triplet extension of the SM,
which shows that our method can be applied also to more realistic cases.

\section*{Acknowledgements}
We thank Antonio Pich and Arcadi Santamaria for comments on the manuscript. 
P.~R. thanks the Instituto de F\'isica Corpuscular (IFIC) in Valencia for hospitality during the completion of this work.
This research has been supported in part by the Spanish Government, by  Generalitat Valenciana and by ERDF funds from the 
EU Commission [grants FPA2011-23778, FPA2014-53631-C2-1-P, PROMETEOII/2013/007, SEV-2014-0398]. J.~F. also acknowledges VLC-CAMPUS for an 
``Atracci\'{o} del Talent"  scholarship.

\appendix
\renewcommand{\theequation}{\Alph{section}.\arabic{equation}}

\section{General expressions for dimension-six operators}
\label{ap:D6formulae}

In this appendix we workout $\mathcal{L}_{\rm EFT}^{\rm 1loop}$ for the case of dimension-six operators. Following the guidelines in Section~\ref{s:method}, we make the separation $U(x,\partial_x+ip)=U_H(x)+U_{LH}(x,\partial_x+ip)$ and expand Eq.~\eqref{eq:L1loop_final} up to $\mathcal{O}(\zeta^{-2})$. The Lagrangian $\mathcal{L}_{\rm EFT}^{\rm 1loop}$ then consists of two pieces: 
\begin{align}\label{eq:1LEFT_expanded}
\mathcal{L}_{\rm EFT}^{\rm 1loop}=\left.\mathcal{L}_{\rm EFT}^{\rm 1loop}\right|_{\rm H}+\left.\mathcal{L}_{\rm EFT}^{\rm 1loop}\right|_{\rm LH}\,.
\end{align}
The first term comes from contributions involving $U_H(x)$ only and, since $U_H(x)$ is momentum independent, it can be obtained from the universal formula provided in the literature~\cite{Ball:1988xg,Bilenky:1993bt,Henning:2014wua} (see also \cite{Drozd:2015rsp} for the case when several scales are involved) which, for completeness, we reproduce here:
\begin{align}\label{eq:Huniversal}
\left.\mathcal{L}_{\rm EFT}^{\rm 1loop}\right|_{\rm H}&=\frac{c_s}{16\pi^2}\left\{m_H^2\,\left(1+\ln\frac{\mu^2}{m_H^2}\right){\rm tr}\left\{U_H\right\}\right. \nonumber\\
&\quad+\left.\Big[\frac{1}{2}\ln\frac{\mu^2}{m_H^2}\,{\rm tr}\left\{U_H^2\right\}+\frac{1}{12}\ln\frac{\mu^2}{m_H^2}\,{\rm tr}\left\{\hat F_{\mu\nu}\hat F^{\mu\nu}\right\}\Big]\right.\nonumber\\
&\quad+\left.\frac{1}{m_H^2}\Big[-\frac{1}{6}{\rm tr}\left\{U_H^3\right\}+\frac{1}{12}{\rm tr}\left\{(\hat D_\mu U_H)^2\right\}-\frac{1}{12}{\rm tr}\left\{U_H \hat F^{\mu\nu}\hat F_{\mu\nu}\right\}\right.\nonumber\\
&\quad\hspace{1.45cm}\left.+\frac{1}{60}{\rm tr}\left\{(\hat D_\mu \hat F^{\mu\nu})^2\right\}-\frac{1}{90}{\rm tr}\left\{\hat F^{\mu\nu}\hat F_{\nu\rho}\hat F^\rho_{\;\;\mu}\right\}\Big]\right.\nonumber\\
&\quad\left.+\,\frac{1}{m_H^4}\Big[\,\frac{1}{24}{\rm tr}\left\{U_H^4\right\}-\frac{1}{12}{\rm tr}\left\{U_H (\hat D_\mu U_H)^2\right\}+\frac{1}{60}{\rm tr}\left\{\hat F_{\mu\nu}(\hat D^\mu U_H)(\hat D^\nu U_H)\right\}\right.\nonumber\\
&\quad\hspace{1.4cm}\left.+\frac{1}{120}{\rm tr}\left\{(\hat D^2 U_H)^2\right\}+\frac{1}{40}{\rm tr}\left\{U_H^2\hat F_{\mu\nu}\hat F^{\mu\nu}\right\}+\frac{1}{60}{\rm tr}\left\{(U_H \hat F_{\mu\nu})^2\right\}\Big]\right.\nonumber\\
&\quad\left.+\frac{1}{m_H^6}\Big[-\frac{1}{60}{\rm tr}\left\{U_H^5\right\}+\frac{1}{20}{\rm tr}\left\{U_H^2(\hat D_\mu U_H)^2\right\}+\frac{1}{30}{\rm tr}\left\{(U_H\hat D_\mu U_H)^2\right\}\Big]\right.\nonumber\\
&\quad+\frac{1}{m_H^8}\frac{1}{120}{\rm tr}\left\{U_H^6\right\}+\mathcal{O}\left(\zeta^{-3}\right)\,,
\end{align}
where $c_s=1/2,-1/2$ depending, respectively, on the bosonic or fermionic nature of the heavy fields. Here $F_{\mu\nu}\equiv\left[D_\mu,D_\nu\right]$ and the momentum integrals are regulated in $d$ dimensions, with the divergences subtracted in the $\overline{\rm MS}$ scheme. The second term in Eq.~\eqref{eq:1LEFT_expanded} is built from pieces containing at least one power of $U_{LH}$. Given that $U_H$ is at most $\mathcal{O}(\zeta)$ and $U_{LH}$ at most $\mathcal{O}(\zeta^0)$ in our power counting, the series in Eq.~\eqref{eq:L1loop_final} has to be expanded up to $n=5$ for the contributions to dimension-six operators
\begin{align}\label{eq:LHuniversal}
\left.\mathcal{L}_{\rm EFT}^{\rm 1loop}\right|_{\rm LH}&=-ic_s\,\int\frac{d^dp}{\left(2\pi\right)^d}\left\{\frac{1}{p^2-m_H^2}\,\mbox{tr}_{\rm s}\left\{U\right\}+\frac{1}{2}\frac{1}{\left(p^2-m_H^2\right)^2}\,\mbox{tr}_{\rm s}\left\{U^2\right\}\right.\nonumber\\
&\quad\left.+\frac{1}{3}\frac{1}{\left(p^2-m_H^2\right)^3}\,\left[\mbox{tr}_{\rm s}\left\{U^3\right\}+\mbox{tr}_{\rm s}\left\{U\hat D^2U\right\}+2ip^\mu\,\mbox{tr}_{\rm s}\left\{U\hat D_\mu U\right\}\right]\right.\nonumber\\
&\quad\left.+\frac{1}{4}\frac{1}{\left(p^2-m_H^2\right)^4}\,\Big[\mbox{tr}_{\rm s}\left\{U^4\right\}+2ip^\mu\,\mbox{tr}_{\rm s}\left\{U^2 \hat D_\mu U\right\}+2ip^\mu\,\mbox{tr}_{\rm s}\left\{U \hat D_\mu U^2\right\}\right.\nonumber\\
&\quad\left.\hspace{3cm}+\,\mbox{tr}_{\rm s}\left\{U^2 \hat D^2 U\right\}+\,\mbox{tr}_{\rm s}\left\{U \hat D^2 U^2\right\}-4\,p^\mu p^\nu\,\mbox{tr}_{\rm s}\left\{U \hat D_\mu \hat D_\nu U\right\}\right.\nonumber\\
&\quad\hspace{3cm}\left.+2ip^\mu\,\mbox{tr}_{\rm s}\left\{U \hat D^2\hat D_\mu U\right\}+2ip^\mu\,\mbox{tr}_{\rm s}\left\{U \hat D_\mu\hat D^2 U\right\}\right.\nonumber\\
&\quad\left.\hspace{3cm}+\mbox{tr}_{\rm s}\left\{U (\hat D^2)^2\, U\right\}\Big]\right.\nonumber\\
&\quad\left.+\frac{1}{5}\frac{1}{\left(p^2-m_H^2\right)^5}\,\Big[\mbox{tr}_{\rm s}\left\{U^5\right\}+2ip^\mu\,\mbox{tr}_{\rm s}\left\{U^3\hat D_\mu U\right\}+2ip^\mu\,\mbox{tr}_{\rm s}\left\{U^2\hat D_\mu U^2\right\}\right.\nonumber\\
&\quad\left.\hspace{3cm}+2ip^\mu\,\mbox{tr}_{\rm s}\left\{U\hat D_\mu U^3\right\}-4p^\mu p^\nu\,\mbox{tr}_{\rm s}\left\{U^2 \hat D_\mu \hat D_\nu U\right\}\right.\nonumber\\
&\quad\left.\hspace{3cm}-4p^\mu p^\nu\,\mbox{tr}_{\rm s}\left\{U \hat D_\mu U \hat D_\nu U\right\}-4p^\mu p^\nu\,\mbox{tr}_{\rm s}\left\{U \hat D_\mu \hat D_\nu U^2\right\}\right.\nonumber\\
&\quad\left.\hspace{3cm}-8i\,p^\mu p^\nu p^\rho\,\mbox{tr}_{\rm s}\left\{U \hat D_\mu \hat D_\nu \hat D_\rho U\right\}\Big]\right\}\nonumber\\
&\quad+\mathcal{L}_{\rm EFT}^F+\mathcal{O}\left(\zeta^{-3}\right)\,.
\end{align}
We have introduced a subtracted trace which is defined as
\begin{align}
\mbox{tr}_{\rm s}\left\{f(U,D_\mu)\right\}\equiv\mbox{tr}\left\{f(U,D_\mu)-f(U_H,D_\mu)-\Theta_f\right\}\,,
\end{align}
where $f$ is an arbitrary function of $U$ and covariant derivatives, and $\Theta_f$ generically denotes all the terms with covariant derivatives at 
the rightmost of the trace ({\em i.e.} open covariant derivative terms) contained in the original trace. The terms involving only $U_H$ that 
are subtracted from the trace were already included in Eq.~\eqref{eq:Huniversal} while all open derivative terms from the different 
traces are collected in $\mathcal{L}_{\rm EFT}^F$. The latter combine into gauge invariant pieces with field-strength tensors, although the manner in which this occurs is not easily seen and involves the contribution from different orders in the expansion. 

With the purpose of illustration, we compute $\mathcal{L}_{\rm EFT}^F$ that results from the integration of the real scalar triplet extension 
of the SM presented in Subsection~\ref{ss:example2}. In this case, gauge invariance of the final result guarantees that the
leading order contribution to $\mathcal{L}_{\rm EFT}^F$ should contain at least four covariant derivatives, as 
terms with two covariant derivatives cannot be contracted to yield a gauge invariant term. As it is clear from Eq.~\eqref{eq:L1loop_final}, 
traces with $j$ derivatives and a number $k$ of $U$ operators have a power suppression of  $\mathcal{O}\left(\zeta^{4-j-2k}\right)$  
(we recall that $d^dp\sim \zeta^4$). The expansion of the operator $U_{LH}$ can yield in addition $\ell$ covariant derivatives, and each of these receives a further suppression of $\zeta^{-1}$  because they are accompanied with a light-field propagator, see Eq.~\eqref{eq:ULH_F}. Since $U_{LH}$ is at most $\mathcal{O}(\zeta^0)$  we then find that, in general, terms with $k$ insertions of $U_{HL}$  and a total number of $j+\ell$ derivatives have a power counting of at most $\mathcal{O}(\zeta^{4-j-\ell-2k})$. As a result, the only gauge invariant object involving $U_{LH}$ and four derivatives that one can construct at $\mathcal{O}(\zeta^{-2})$ includes only one power of $U_{LH}$ ({\em i.e.} $j+\ell=4$ and $k=1$).
Moreover, since $U_{LH}$ has to be evaluated at leading order, the only relevant piece from $U_{LH}$ for the computation of $\mathcal{L}_{\rm EFT}^F$ reads 
\begin{align}
U_{LH}^F=X_{LH}^{(1)\,\dagger}\left.\Delta_L^{-1}\right|_{\hat\eta=0}X_{LH}^{(1)}\,.
\end{align}
Here $X_{LH}^{(1)}$ is defined as the part of $X_{LH}$ that is $\mathcal{O}\left(\zeta\right)$, and we remind that $\hat\eta$ stands for the classical field configurations. Using the expressions in Eqs.~\eqref{eq:fluctuatri} and~\eqref{eq:fluctus} we have 
\begin{align}\label{eq:ULH_F}
U_{LH}^F(x,\partial_x+ip)&\subset \frac{\kappa^2}{p^2}\sum_{m=0}^4\left[\,\hat\phi^\dagger \tau^a \left(\frac{2ip \hat D+\hat D^2}{p^2}\right)^m \tau^b\hat\phi+\hat\phi^\intercal (\tau^a)^\intercal\left(\frac{2ip \hat D^*+\hat D^{*\,2}}{p^2}\right)^m (\tau^{b})^*\hat\phi^*\right]\,,
\end{align}
where the covariant derivatives have to be expanded by applying the identities
\begin{align}
\begin{aligned}
D_\mu \tau^a \phi&=\tau^a \left(D_\mu\phi\right)+\tau^c\phi\, D_\mu^{ca}\,,\\
D^*_\mu\, (\tau^{a})^* \phi^*&=(\tau^a)^* \left(D_\mu\phi\right)^*+(\tau^c)^*\phi^*\, D_\mu^{ca}\,,
\end{aligned}
\end{align}
with $D_\mu$ denoting the Higgs field covariant derivative, see Eq.~\eqref{eq:SM_cdev},
 and with $D_\mu^{ca}$ as defined in Section~\ref{s:method}. For the computation of $\mathcal{L}_{\rm EFT}^F$ up to $\mathcal{O}\left(\zeta^{-2}\right)$ we need to isolate the terms in Eq.~\eqref{eq:L1loop_final} with up to four open covariant derivatives and just one power of $U_{LH}^F$. These are given by
\begin{align}
\begin{aligned}
\mathcal{L}_{\rm EFT}^F&\subset
-\frac{i}{2}\int\frac{d^d p}{\left(2\pi\right)^d}\frac{1}{p^2-M^2}\sum_{n=0}^4\sum_{k=0}^n \frac{1}{n+1}\\
&\quad\times{\rm tr}\left\{\left(\frac{2ip \hat D+\hat D^2}{p^2-M^2}\right)^{n-k} U^F_{LH}(x,\partial_x+ip) \left(\frac{2ip \hat D+\hat D^2}{p^2-M^2}\right)^k\right\}\,,
\end{aligned}
\end{align}
and using the cyclic property of the trace we get\footnote{The use of the cyclic property when derivative terms are involved is only justified for the functional trace, that we denoted in this article as ${\rm Tr}$. 
However, as noted in Refs.~\cite{Chan:1986jq,Ball:1988xg}, in the evaluation of the functional determinant, which is a gauge invariant object, the trace over internal degrees of freedom `${\rm tr}$' can be recast 
into the full trace through the use of the identity (we recall that $S=\int d^dx \,\mathcal{L}$)
\begin{align*}
{\rm Tr} \{f(\hat x)\}=\int d^dx \;{\rm tr}\{\langle x| f(\hat x)|x\rangle\}=\int d^dx \;{\rm tr}\{f(x)\}\,\delta^d(0)\,,
\end{align*}
and then reverted to a trace over internal degrees of freedom after the application of the cyclic property.}
\begin{align}\label{eq:LF_general}
\mathcal{L}_{\rm EFT}^F&\subset
-\frac{i}{2}\int\frac{d^d p}{\left(2\pi\right)^d}\frac{1}{p^2-M^2}\sum_{n=0}^4\,{\rm tr}\left\{U^F_{LH}(x,\partial_x+ip) \left(\frac{2ip \hat D+\hat D^2}{p^2-M^2}\right)^n\right\}\,.
\end{align}
Finally, keeping only terms with up to four covariant derivatives, performing the momentum integration (see Eq.~\eqref{eq:master_int}) and evaluating the $SU(2)$ trace we arrive at the final result
\begin{align}
\begin{aligned}
\mathcal{L}_{\rm EFT}^F&=\frac{1}{16\pi^2}\frac{\kappa^2}{M^4}\left[-\frac{g^2}{3}\,\left(\hat\phi^\dagger\hat\phi\right)\, \hat W^{\mu\nu\,a}\,\hat W^a_{\mu\nu}+g\, \big(\hat\phi^\dagger\,i\lra{\hat{D}_\mu^a}\,\hat\phi\big)\big(\hat D_\nu\,\hat W^{\mu\nu}\big)^a-\frac{g g^\prime}{2}\left(\hat\phi^\dagger\tau^a\hat\phi\right)\hat W_{\mu\nu}^a\hat B^{\mu\nu}\right]\,,
\end{aligned}
\end{align}
with the field-strength tensors defined in Eq.~\eqref{eq:strength_tensors} and
\begin{align}
\big(\phi^\dagger\,i\lra{D^a_\mu}\phi\big)=i\,\big(\phi^\dagger\,\tau^a D_\mu\phi\big)-i\big[(D_\mu\phi)^\dagger\,\tau^a\phi\big]\,.
\end{align}

\section{The fluctuation operator of the SM}
\label{ap:SMfluc}
\setcounter{equation}{0}

In this appendix we provide the fluctuation operator for the SM Lagrangian. The SM Lagrangian in compact notation is given by
\begin{align}\label{eq:SM_lag}
\begin{aligned}
\mathcal{L}_{\mbox{\scriptsize SM}}=&-\frac{1}{4}G_{\mu\nu}^\alpha G^{\mu\nu\,\alpha}-\frac{1}{4}W_{\mu\nu}^aW^{\mu\nu\,a}-\frac{1}{4}B_{\mu\nu}B^{\mu\nu}
+\left(D_\mu\phi\right)^\dagger D^\mu\phi-m_\phi^2\left(\phi^\dagger\phi\right)-\frac{\lambda}{2}\left(\phi^\dagger\phi\right)^2\\
&+\overline\psi\,i\slashed{D}\,\psi-\overline\psi\left(\widetilde\phi\,y_u\, P_uP_R+\phi\,y_d\, P_dP_R+h.c.\right)\psi
+\mathcal{L}_{\mbox{\scriptsize GF}}+\mathcal{L}_{\mbox{\scriptsize ghost}}\,.
\end{aligned}
\end{align}
Here, $\psi=q,\ell$, $P_u\,(P_d)$ project into the up (down) sector, $y_{u,d}$ is a Yukawa matrix for up (down) fields, $\mathcal{L}_{\mbox{\scriptsize GF}}$ and $\mathcal{L}_{\mbox{\scriptsize ghost}}$ are the gauge-fixing and ghost Lagrangians, respectively, and the covariant derivatives are defined as
\begin{align}\label{eq:SM_cdev}
\begin{aligned}
\slashed{D}\psi&=\left(\slashed{\partial}_\mu-ig_c\slashed{G}^\alpha T^\alpha P_q-ig\slashed{W}^a T^aP_L-ig^\prime \slashed{B} Y_\psi\right)\psi,\\
D_\mu\phi&=\left(\partial_\mu-igW_\mu^a T^a-\frac{1}{2}ig^\prime B_\mu\right)\phi \,.
\end{aligned}
\end{align}
In Eq.~(\ref{eq:SM_cdev}), $T^a=\tau^a/2$ and  $T^\alpha=\lambda^\alpha/2$ with $\tau^a$ and $\lambda^\alpha$ the Pauli and the Gell-Mann matrices, respectively, $P_q$ denotes a projector into the quark sector, and the hypercharge reads $Y_\psi=Y_{\psi_L} P_L+Y^u_{\psi_R} P_u P_R+Y^d_{\psi_R} P_d P_R$. Accordingly, the field strength tensors are given by
\begin{align}\label{eq:strength_tensors}
\begin{aligned}
G_{\mu\nu}^\alpha&=\partial_\mu G_\nu^\alpha-\partial_\nu G_\mu^\alpha+g f_{\alpha\beta\gamma}G_\mu^\beta\,G_\nu^\gamma,\\
W_{\mu\nu}^a&=\partial_\mu W_\nu^a-\partial_\nu W_\mu^a+g\epsilon_{abc}W_\mu^b\,W_\nu^c,\\
B_{\mu\nu}&=\partial_\mu B_\nu-\partial_\nu B_\mu.
\end{aligned}
\end{align}

Following the same procedure as in Section~\ref{s:method}, we separate the fields into 
background, $\hat \eta$, and quantum field configurations, $\eta$, and expand the SM Lagrangian to second order in the quantum fluctuation:
\begin{align}\label{eq:SM_eff_Lag}
\mathcal{L}_{\mbox{\scriptsize SM}}=\mathcal{L}_{\mbox{\scriptsize SM}}^{\mbox{\scriptsize tree}}(\hat{\eta})+\mathcal{L}_{\mbox{\scriptsize SM}}^{(\eta^2)}+\mathcal{O}\left(\eta^3\right)\,,
\end{align}
where $\mathcal{L}_{\mbox{\scriptsize SM}}^{\mbox{\scriptsize tree}}$ is the tree-level SM effective Lagrangian, and $\mathcal{L}_{\mbox{\scriptsize SM}}^{(\eta^2)}$ is computed using Eq.~\eqref{eq:2ndorderL}:
\begin{align}\label{eq:SM_1loop}
\mathcal{L}_{\mbox{\scriptsize SM}}^{(\eta^2)}&=\frac{1}{2}
\setlength\arraycolsep{1.6pt}
\begin{pmatrix}
\phi^\dagger & \phi^\intercal & A_\mu^{a\,\intercal} & \overline{\psi} & \psi^\intercal
\end{pmatrix}
\setlength\arraycolsep{4pt}
\begin{pmatrix}
\Delta_{\phi^*\phi} & X_{\phi\phi}^\dagger & \left(X_{A\phi}^{\nu\,b}\right)^\dagger & \overline X_{\bar \psi \phi} & -X_{\bar \psi \phi^*}^\intercal\\[5pt]
X_{\phi\phi} & \Delta_{\phi^*\phi}^\intercal & \left(X_{A\phi}^{\nu\,b}\right)^\intercal & \overline X_{\bar{\psi}\phi^*} & -X_{\bar{\psi}\phi}^\intercal\\[5pt]
X_{A\phi}^{\mu\,a} & \left(X_{A\phi}^{\mu\,a}\right)^* & \Delta_A^{\mu\nu\,ab} & \overline X_{\bar \psi A}^{\mu\,a} & -\left(X_{\bar \psi A}^{\mu\,a}\right)^\intercal\\[5pt]
X_{\bar \psi \phi} & X_{\bar{\psi}\phi^*} & X_{\bar \psi A}^{\nu\,b} & \Delta_{\bar\psi\psi} & 0\\[5pt]
-\overline X_{\bar \psi \phi^*}^\intercal & -\overline X_{\bar{\psi}\phi}^\intercal & -\big(\overline X_{\bar \psi A}^{\,\nu\,b}\big)^\intercal & 0 & -\Delta_{\bar\psi\psi}^\intercal &\\[5pt]
\end{pmatrix}
\begin{pmatrix}
\phi\\[5pt]
\phi^*\\[5pt]
A_\nu^b\\[5pt]
\psi\\[5pt]
\overline{\psi}^\intercal
\end{pmatrix}
+\mathcal{L}_{\mbox{\scriptsize ghost}}^{(\eta^2)},
\end{align}
with $A_\mu^a=\left(G_\mu^\alpha\,\;W_\mu^a\,\;B_\mu\right)^\intercal$ denoting the gauge fields and
\begin{align}
\Delta_A^{\mu\nu\,ab}=
\begin{pmatrix}
\Delta_G^{\mu\nu\,\alpha\beta} & 0 & 0\\[5pt]
0 & \Delta_W^{\mu\nu\,ab} & \Delta_{BW}^{\mu\nu\,a}\\[5pt]
0 & \Delta_{BW}^{\mu\nu\,a} & \Delta_B^{\mu\nu}
\end{pmatrix}
\quad,\quad 
X_{A\phi}^{\mu\,a}=
\begin{pmatrix}
0\\[5pt]
X_{W\phi}^{\mu\,a}\\[5pt]
X_{B\phi}^\mu
\end{pmatrix}
\quad,\quad 
\overline X_{\bar \psi A}^{\mu\,a}=
\begin{pmatrix}
\overline X_{\bar \psi G}^{\mu\,\alpha}\\[5pt]
\overline X_{\bar \psi W}^{\mu\,a}\\[5pt]
\overline X_{\bar \psi B}^\mu
\end{pmatrix}
\,,
\end{align}
where, generically, $\overline{X}=X^\dagger \gamma^0$. The pieces in the quadratic fluctuation are defined as
\begin{align}\label{eq:SM_fluctuations}
\Delta_{\phi^*\phi}=&-\hat{D}^2-m_\phi^2-\lambda\left(\hat{\phi}^\dagger\hat{\phi}\right)-\lambda\hat{\phi}\hat{\phi}^\dagger,\nonumber\\
\Delta_G^{\mu\nu\,\alpha\beta}=&\,\delta_{\alpha\beta}\left[g^{\mu\nu}\hat{D}^2+\frac{1-\xi_G}{\xi_G}\hat{D}^\mu \hat{D}^\nu\right]-g_c\epsilon_{\alpha\beta\gamma}\hat{G}^{\mu\nu\,\gamma},\nonumber\\
\Delta_W^{\mu\nu\,ab}=&\,\delta_{ab}\left\{g^{\mu\nu}\left[\hat{D}^2+\frac{1}{2}g^2\left(\hat{\phi}^\dagger\hat{\phi}\right)\right]+\frac{1-\xi_W}{\xi_W}\hat{D}^\mu \hat{D}^\nu\right\}
-g\epsilon_{abc}\hat{W}^{\mu\nu\,c},\nonumber\\
\Delta_B^{\mu\nu}=&\,g^{\mu\nu}\left[\partial^2+\frac{1}{2}g^{\prime\,2}\left(\hat{\phi}^\dagger\hat{\phi}\right)\right]+\frac{1-\xi_B}{\xi_B}\partial^\mu \partial^\nu,\nonumber\\
\Delta_{BW}^{\mu\nu\,a}=&\,\frac{1}{2}gg^\prime g_{\mu\nu}\left(\hat{\phi}^\dagger \tau^a\hat{\phi}\right),\nonumber\\
\Delta_{\bar\psi\psi}=&\,i\hat{\slashed{D}}-\left(i\tau_2\,\hat \phi^*\,y_u\, P_u+\hat\phi\,y_d\, P_d+h.c.\right),\nonumber\\
X_{\phi\phi}=&-\lambda\,\hat{\phi}^*\hat{\phi}^\dagger\,,\nonumber\\
X_{W\phi}^{\mu\,a}=&\,\frac{1}{2}ig\left[\hat{\phi}^\dagger\tau^a\hat{D}^\mu-\left(\hat D^\mu\hat\phi\right)^\dagger\tau^a\right],\nonumber\\
X_{B\phi}^\mu=&\,\frac{1}{2}ig^\prime\left[\hat{\phi}^\dagger\hat{D}^\mu-\left(\hat{D}^\mu\hat{\phi}\right)^\dagger\right],\nonumber\\
X_{\bar \psi G}^{\mu\,\alpha}=&\,\frac{1}{2}g_c\, \lambda^\alpha P_q\,\gamma^\mu\hat{\psi}\,,\nonumber\\
X_{\bar \psi W}^{\mu\,a}=&\,\frac{1}{2}g\, \tau^a\,\gamma^\mu P_L\hat{\psi}\,,\nonumber\\
X_{\bar \psi B}^\mu=&\,g^\prime \gamma^\mu Y_\psi\,\hat{\psi}\,,\nonumber\\
X_{\bar \psi \phi}=&\,-\la{P}_uP_L y_u^\dagger\,\hat\psi^t\,i\tau_2-y_d\,P_dP_R\,\hat\psi\,,\nonumber\\
X_{\bar{\psi}\phi^*}=&\,-i\tau^2\,y_u P_uP_R \hat\psi-\la{P}_dP_L y_d^\dagger\hat\psi^t\,.
\end{align}
The superscript $t$ in the fermion fields denotes transposition in isospin space. 
Additionally, we have fixed the gauge of the quantum fields using the \textit{background field gauge},
which ensures that the theory remains invariant under gauge transformations of the background fields. This choice corresponds to the following gauge-fixing Lagrangian:
\begin{align}\label{eq:ghost_Lag}
\begin{aligned}
\mathcal{L}_{\mbox{\scriptsize GF}}=&-\frac{1}{2\xi_G}\left(\hat{D}_\mu G^{\mu\,\alpha}\right)^2-\frac{1}{2\xi_W}\left(\hat{D}_\mu W^{\mu\,a}\right)^2
-\frac{1}{2\xi_B}\left(\partial_\mu B^\mu\right)^2\,.
\end{aligned}
\end{align}

Finally we also provide the expansion for the inverse operators $\Delta_X\left(x,\partial_x+ip\right)^{-1}$, with $X=\left\{\phi^*\phi,\,B,\,W\right\}$, when $p_\mu\sim\zeta$. We have:
\begin{align}\label{eq:SM_Delta_fluc}
\begin{aligned}
\Delta_{\phi^*\phi}\left(x,\partial_x+ip\right)&=p^2-m_\phi^2-2ip\hat{D}-\hat{D}^2-\lambda\left(\hat{\phi}^\dagger\hat{\phi}\right)
-\lambda\hat{\phi}\hat{\phi}^\dagger\,,\\
\Delta_W^{\mu\nu\,ab}\left(x,\partial_x+ip\right)&=\delta_{ab}\left[-g^{\mu\nu}p^2-\frac{1-\xi_W}{\xi_W}p^\mu p^\nu\right]
+\mathcal{O}\left(\zeta\right)\,,\\
\Delta_B^{\mu\nu}\left(x,\partial_x+ip\right)&=-g^{\mu\nu}p^2-\frac{1-\xi_B}{\xi_B}p^\mu p^\nu+\mathcal{O}\left(\zeta\right)\,.
\end{aligned}
\end{align}
from where, and defining
\begin{equation} \label{eq:omegaux}
\Omega \, = \, \hat{D}^2+\lambda\left(\hat{\phi}^\dagger\hat{\phi}\right)+\lambda\,\hat{\phi}\hat{\phi}^\dagger \, ,
\end{equation}
it is straightforward to get
\begin{align}\label{eq:SM_inv_prop}
\begin{aligned}
\Delta_{\phi^*\phi}\left(x,\partial_x+ip\right)^{-1}&=\,\frac{1}{p^2}\left(1+\frac{m_\phi^2}{p^2}+\frac{m_\phi^4}{p^4}\right)+ 2 i \, \frac{p_\mu}{p^4}\left(1+2\frac{m_\phi^2}{p^2}\right)\, \hat{D}^\mu\\
&\quad+\frac{1}{p^4}\left(1+2\frac{m_\phi^2}{p^2}\right) \, \Omega \, 
- 4 \, \frac{p_\mu p_\nu}{p^6}\left(1+3\frac{m_\phi^2}{p^2}\right)\,\hat{D}^\mu\hat{D}^\nu\\
&\quad+2 i \, \frac{p_\mu}{p^6}\,\left\{\hat{D}^\mu \, \Omega \, + \, \Omega \, \hat{D}^\mu\right\} \, + \, \frac{1}{p^6} \, \Omega^2\\
&\quad-8i \,  \frac{p_\mu p_\nu p_\rho}{p^8}\,\hat{D}^\mu\hat{D}^\nu\hat{D}^\rho \,
+ 16 \,  \frac{p_{\mu} p_{\nu} p_{\rho} p_{\sigma}}{p^{10}} \, \hat{D}^{\mu} \hat{D}^{\nu} \hat{D}^{\rho} \hat{D}^{\sigma} \\
&\quad-4 \, \frac{p_\mu  p_\nu}{p^8}\left\{\hat{D}^\mu\hat{D}^\nu \, \Omega \, + \, \Omega \, \hat{D}^\mu\hat{D}^\nu +\,\hat{D}^\mu \, \Omega \, \hat{D}^\nu\right\} +\mathcal{O}\left(\zeta^{-7}\right),\\
\Delta_B^{\mu\nu}\left(x,\partial_x+ip\right)^{-1}&=-\frac{g^{\mu\nu}}{p^2}+\left(1-\xi_B\right)\frac{p^\mu p^\nu}{p^4}+\mathcal{O}\left(\zeta^{-3}\right)\,,\\
\Delta_W^{\mu\nu\,ab}\left(x,\partial_x+ip\right)^{-1}&=\delta_{ab}\left[-\frac{g^{\mu\nu}}{p^2}+\left(1-\xi_W\right)\frac{p^\mu p^\nu}{p^4}\right]+\mathcal{O}\left(\zeta^{-3}\right)\,,
\end{aligned}
\end{align}
and analogously for $\Delta_G^{\mu\nu\,\alpha \beta}\left(x,\partial_x+ip\right)^{-1}$. The inverse operator $[\Delta_{\phi^*\phi}^*\left(x,\partial_x+ip\right)]^{-1}$ can be obtained from $\Delta_{\phi^*\phi}\left(x,\partial_x+ip\right)^{-1}$ by making the substitution $\hat D_\mu\to \hat D_\mu^*$ while $[\Delta_{\phi^*\phi}^\intercal\left(x,\partial_x+ip\right)]^{-1}$ and $[\Delta_{\phi^*\phi}^*\left(x,\partial_x+ip\right)]^{-1}$ share the same expression, up to a total derivative term.

\end{document}